\documentclass[aps,prc,reprint,showpacs,superscriptaddress]{revtex4-1}

\usepackage{graphicx,color,rotating,pifont}
\usepackage{amsmath,amssymb,bm}
\usepackage{ae}
\usepackage{pstricks}
\usepackage{dcolumn}

\DeclareMathSymbol{\NS}{\mathord}{AMSb}{"4E}

\newcommand{\ket}[1]{\ensuremath{\,|{#1}\rangle}}

\newcommand{\matrixe}[3]{\ensuremath{\langle{#1}|\,{#2}\,|{#3}\rangle}}

\newcommand{\comm}[2]{\ensuremath{[{#1},{#2}]}}

\newcommand{\op}[1]{\ensuremath{#1}}
\newcommand{\adj}[1]{\ensuremath{{{#1}}^{\dag}}}

\newcommand{\sixj}[1]{
 \ensuremath{
   \begin{Bmatrix}
     #1
   \end{Bmatrix}
 }
}

\newcommand{\aO}{\ensuremath{\op{a}}}

\newcommand{\etaO}{\ensuremath{\op{\eta}}}

\newcommand{\aaO}{\ensuremath{\adj{\op{a}}}}

\newcommand{\HO}{\ensuremath{\op{H}}}

\newcommand{\OO}{\ensuremath{\op{O}}}

\newcommand{\UO}{\ensuremath{\op{U}}}

\newcommand{\hJ}{\widehat{J}}

\newcommand{\totd}[2]{\ensuremath{ \frac{d {#1}} {d {#2}} }}

\newcommand{\eMax}{\ensuremath{e_{\text{Max}}}}

\newcommand{\EMax}{\ensuremath{E_{3\text{Max}}}}

\newcommand{\nuc}[2]{\ensuremath{{}^{#2}\mathrm{#1}}}

\newcommand{\fm}{\ensuremath{\,\text{fm}}}

\newcommand{\keV}{\ensuremath{\,\text{keV}}}
\newcommand{\MeV}{\ensuremath{\,\text{MeV}}}

\newcommand{\symboldiamond}[1][black]{{\color{#1}$\blacklozenge$}}
\newcommand{\symboltriangle}[1][black]{{\color{#1}$\blacktriangle$}}
\newcommand{\symbolbox}[1][black]{{\color{#1}$\blacksquare$}}
\newcommand{\symbolcircle}[1][black]{{\color{#1}\ding{108}}}

\definecolor{FGViolet}{rgb}{0.61,0.32,0.61}
\definecolor{FGDarkBlue}{rgb}{0,0,0.6}
\definecolor{FGBlue}{rgb}{0,0,0.8}
\definecolor{FGLightBlue}{rgb}{0.2, 0.6, 0.8}
\definecolor{FGGreen}{rgb}{0.2,0.7,0.2}
\definecolor{FGLightGreen}{rgb}{0.4,1,0.4}
\definecolor{FGYellow}{rgb}{1,0.95,0}
\definecolor{FGOrange}{rgb}{0.95,0.5,0.1}
\definecolor{FGRed}{rgb}{0.8,0,0}
\definecolor{FGWhite}{rgb}{1,1,1}
\definecolor{FGLightGray}{rgb}{0.8,0.8,0.8}
\definecolor{FGGray}{rgb}{0.5,0.5,0.5}
\definecolor{FGDarkGray}{rgb}{0.3,0.3,0.3}
\definecolor{FGBlack}{rgb}{0,0,0}

\begin{document}
\title{In-Medium Similarity Renormalization Group with Chiral Two- Plus Three-Nucleon Interactions}

\author{H. Hergert}
\affiliation{The Ohio State University, Columbus, OH 43210, USA}
\email{Corresponding author. Electronic address: hergert.3@osu.edu}

\author{S. K. Bogner}
\affiliation{National Superconducting Cyclotron Laboratory and Dept. of Physics and Astronomy, Michigan State University, East Lansing, MI 48824, USA}
\author{S. Binder}
\author{A. Calci}
\author{J. Langhammer}
\author{R. Roth}
\affiliation{Institut f\"ur Kernphysik, Technische Universit\"at Darmstadt,
D-64289 Darmstadt, Germany}
\author{A. Schwenk}
\affiliation{ExtreMe Matter Institute, GSI Helmholtzzentrum f\"ur Schwerionenforschung GmbH, D-64291 Darmstadt, Germany}
\affiliation{Institut f\"ur Kernphysik, Technische Universit\"at Darmstadt,
D-64289 Darmstadt, Germany}

\date{\today}

\begin{abstract}
We use the recently proposed In-Medium Similarity Renormalization Group (IM-SRG) to carry out a systematic study of closed-shell nuclei up to $\nuc{Ni}{56}$, based on chiral two- plus three-nucleon interactions. We analyze the capabilities of the IM-SRG by comparing our results for the ground-state energy to Coupled Cluster calculations, as well as to quasi-exact results from the Importance-Truncated No-Core Shell Model. Using chiral two- plus three-nucleon Hamiltonians whose resolution scales are lowered by free-space SRG evolution, we obtain good agreement with experimental binding energies in $\nuc{He}{4}$ and the closed-shell oxygen isotopes, while the calcium and nickel isotopes are somewhat overbound. 
\end{abstract}


\maketitle

\clearpage

\section{Introduction\label{sec:intro}}
The adaptation of Renormalization Group (RG) and Effective Field Theory (EFT) methods has led to many advances in the treatment of the nuclear many-body problem in recent years \cite{Bogner:2010pq,Epelbaum:2009ve}. The Similarity Renormalization Group (SRG), in particular, has become the tool of choice to lower the resolution scale of two- ($NN$) and three-nucleon ($3N$) interactions \cite{Bogner:2007od,Jurgenson:2009bs,Hebeler:2012ly}. The improved convergence properties of the resulting effective interactions make them an ideal input for many \emph{ab initio} methods. As a consequence, the range of nuclei to which such methods can be applied successfully has increased significantly.

Recently, the In-Medium SRG (IM-SRG) has been proposed as a new \emph{ab initio} many-body technique, which uses the powerful SRG flow-equation framework to directly calculate nuclear structure observables, or to derive effective Hamiltonians for Shell Model calculations, which can provide complete spectroscopic information for closed- and open-shell nuclei from first principles \cite{Tsukiyama:2011uq,Tsukiyama:2012fk}. Here, we use the natural capabilities of the IM-SRG to include $3N$ interactions in normal-ordered two-body approximation, and present a systematic study of closed-shell nuclei up to $\nuc{Ni}{56}$, based on chiral $NN$ and $3N$ interactions.

In Sect.~\ref{sec:formalism}, we give an overview of the IM-SRG formalism, and discuss its connection to Many-Body Perturbation Theory (MBPT). Section \ref{sec:ham_implement} describes the Hamiltonians used in this work and provides some technical details on our calculations. In Sect.~\ref{sec:results}, we compare IM-SRG ground-state energies with results from the non-perturbative Coupled Cluster (CC) method at different truncation levels (see, e.g. \cite{Shavitt:2009}), as well as the Importance-Truncated No-Core Shell Model (IT-NCSM) \cite{Roth:2007fk,Roth:2009eu}. The IT-NCSM approach is a diagonalization of the Hamiltonian in a Hilbert space spanned by importance-sampled basis states. Aside from small uncertainties associated with the importance sampling and model space extrapolations, the IT-NCSM results are exact, and particularly useful for quantifying the uncertainties of our IM-SRG results for closed-shell nuclei, which are presented at the end of Sect.~\ref{sec:results}.  

\section{\label{sec:formalism}Formalism}
\subsection{\label{sec:nord}Normal Ordering}
Our starting point is an intrinsic nuclear $A$-body Hamiltonian containing both $NN$ and $3N$ interactions,
\begin{equation}
  H = \left(1-\frac{1}{A}\right)T + T^{(2)} + V^{(2)} +V^{(3)}\,,
\end{equation}
where $T^{(2)}$ is the two-body part of the intrinsic kinetic energy (see, e.g., \cite{Hergert:2009wh}). Choosing a single Slater determinant $\ket{\Phi}$ as the reference state, we can rewrite the Hamiltonian in terms of normal-ordered operators,
\begin{align}\label{eq:Hno}
  \HO &= E + \sum_{12}f_{12}:\aaO_{1}\aO_{2}: + \frac{1}{4}\sum_{1234}\Gamma_{1234}:\aaO_{1}\aaO_{2}\aO_{4}\aO_{3}:\notag\\
   	  &\hphantom{=}+ \frac{1}{36}\sum_{123456}W_{123456}:\aaO_{1}\aaO_{2}\aaO_{3}\aO_{6}\aO_{5}\aO_{4}:\,,
\end{align}
where the indices include all single-particle quantum numbers, and the strings of creation and annihilation operators obey
\begin{equation}
  \matrixe{\Phi}{:\aaO_{1}\ldots\aO_{2}:}{\Phi} = 0\,.
\end{equation}
It is convenient to work in the eigenbasis of the one-body density matrix in the following, so that
\begin{equation}
  \rho_{ab}=n_{a}\delta_{ab}\,,\quad n_{a}\in\{0,1\}\,,
\end{equation}
and the individual normal-ordered contributions in Eq.~\eqref{eq:Hno} are
\begin{align}
  E &= \left(1-\frac{1}{A}\right)\sum_{a}\matrixe{a}{T}{a}n_{a}\notag\\
  		&\hphantom{=}+ \frac{1}{2}\sum_{ab}\matrixe{ab}{T^{(2)}\!+\!V^{(2)}}{ab}n_{a}n_{b}\notag\\
  		&\hphantom{=}+ \frac{1}{6}\sum_{abc}\matrixe{abc}{V^{(3)}}{abc}n_{a}n_{b}n_{c}\,,\label{eq:E0}\\
  f_{12} &= \left(1-\frac{1}{A}\right)\matrixe{1}{T}{2} 
  		+ \sum_{a}\matrixe{1a}{T^{(2)}\!+\!V^{(2)}}{2a}n_{a}\notag\\
  		&\hphantom{=}+ \frac{1}{2}\sum_{ab}\matrixe{1ab}{V^{(3)}}{2ab}n_{a}n_{b}\,,\label{eq:f}		\\
  \Gamma_{1234} &= \matrixe{12}{T^{(2)}\!+\!V^{(2)}}{34} + \sum_{a}\matrixe{12a}{V^{(3)}}{34a}n_{a}\,,\label{eq:Gamma}\\
  W_{123456}&=\matrixe{123}{V^{(3)}}{456}\,.
\end{align}
Due to the occupation numbers in Eqs.~\eqref{eq:E0}--\eqref{eq:Gamma}, the sums run over occupied (hole) states only. Note that the zero-, one-, and two-body parts of the Hamiltonian all contain in-medium contributions from the free-space $3N$ interaction. The normal-ordered $3N$ contribution $W$ will be omitted in the following, leading to the normal-ordered two-body approximation (NO2B), which has been shown to be a very good approximation for the nuclei considered in this work \cite{Roth:2012qf,Binder:2012uq,Hagen:2007zc}. 
 
\subsection{IM-SRG Flow Equations}
The aim of the IM-SRG is to decouple the ground-state of the Hamiltonian from all excitations by means of a continuous unitary transformation. The transformed Hamiltonian is defined as 
\begin{equation}
  H(s)=U^{\dag}(s)H(0)U(s)\,,
\end{equation}
which, upon taking the derivative w.r.t. the flow parameter $s$, yields the following first-order operator differential equation:
\begin{equation}\label{eq:flowabs}
  \totd{}{s}H(s)=\comm{\eta(s)}{H(s)}\,,
\end{equation}
with the generator formally defined by 
\begin{equation}
  \eta(s)=\totd{U^{\dag}(s)}{s}U(s) = -\eta^{\dag}(s)\,.
\end{equation}

When carried out exactly, the IM-SRG is a unitary transformation in $A$-nucleon space, and consequently, $\eta(s)$ and $\HO(s)$ are $A$-body operators. When Eq.~\eqref{eq:flowabs} is integrated, every evaluation of the commutator increases the particle rank of $\HO(s)$, e.g.,
\begin{equation}
  \comm{:\aaO_{1}\aaO_{2}\aO_{4}\aO_{3}:}{:\aaO_{5}\aaO_{6}\aO_{8}\aO_{7}:}= \delta_{35}:\aaO_{1}\aaO_{2}\aaO_{6}\aO_{8}\aO_{7}\aO_{4}:+\ldots.
\end{equation}
All of these induced contributions will in turn contribute to the parts of $\HO(s)$ with lower particle rank in subsequent integration steps. Because an explicit treatment of all contributions up to the $A$-body level is clearly not feasible, we have to introduce a truncation to close the system of IM-SRG flow equations. The simplest approach is to truncate $\HO(s)$ at a given particle rank $n\leq A$, which is motivated by the cluster decomposition principle for short-range interactions (see, e.g., \cite{Weinberg:1996uf}). It has been shown that the omission of $W$, the residual $3N$ part of the Hamiltonian \eqref{eq:Hno}, is a good approximation as long as the in-medium contributions of the free-space $3N$ interaction are accounted for by the normal-ordered zero-, one-, and two-body interactions \cite{Roth:2012qf,Binder:2012uq,Hagen:2007zc}. We therefore truncate both $\HO(s)$ and $\eta(s)$ at the two-body level, and refer to this truncation as IM-SRG(2). An additional truncation which was motivated by a perturbative analysis was proposed in \cite{Tsukiyama:2011uq}, but will not be considered in this work. For the Hamiltonians considered in the following, the numerical results of the two truncation schemes agree to within a few keV \cite{Tsukiyama:2011uq}.

Note that the presence of the commutator in the flow equation \eqref{eq:flowabs} guarantees the IM-SRG to be size-extensive, i.e., the IM-SRG wave-function $\UO(s)\ket{\Phi}$ can be expanded in terms of linked diagrams only \cite{Brandow:1967tg,Bartlett:1981zr,Negele:1998ve}. As a consequence, errors which are introduced by truncating the many-body expansion scale linearly with $A$ \cite{Brandow:1967tg,Negele:1998ve}, and we can, in principle, judge the quality of the IM-SRG results in medium- or heavy-mass nuclei based on comparisons with results from exact methods like the No-Core Shell Model (NCSM) \cite{Navratil:2000nm,Navratil:2009hc}, which are only available for light nuclei.

Since the focus of this work is on closed-shell nuclei, we assume spherical symmetry. Single-particle indices collectively represent the radial, angular momentum, and isospin quantum numbers $i=(k_{i}l_{i}j_{i}\tau_{i})$, and do not depend on the angular momentum projection $m_{i}$. The matrix elements of single-particle operators are diagonal in all but the radial quantum numbers, e.g.,
\begin{equation}
  f_{12}=f^{l_{1}j_{1}\tau_{1}}_{k_{1}k_{2}}\delta_{l_{1}l_{2}}\delta_{j_{1}j_{2}}\delta_{\tau_{1}\tau_{2}}\,,
\end{equation}
and two-body matrix elements are coupled to good $J$ and independent of $M$. Suppressing all $s$-dependence for brevity, the resulting $J$-scheme flow equations read:
\begin{widetext}
\begin{align}
  \label{eq:0bflow}
  \totd{E}{s}&=\sum_{ab}\hat{j}_{a}^{2}\eta_{ab}f_{ba}(n_{a}-n_{b})
					+\frac{1}{2}\sum_{abcdJ}\hJ^{2}\eta^{J}_{abcd}\Gamma^{J}_{cdab}n_{a}n_{b}(1-n_{c})(1-n_{d})\,,\\
  \label{eq:1bflow}
  \totd{f_{12}}{s}&=\sum_{a}\eta_{1a}f_{a2}-f_{1a}\eta_{a2}
  				 	+\frac{1}{\hat{j}^{2}_{1}}\sum_{abJ}\hJ^{2}(n_{a}-n_{b})\left(\eta_{ab}\Gamma^{J}_{b1a2}-f_{ab}\eta^{J}_{b1a2}\right)\notag\\                 
  				 &\hphantom{=}+\frac{1}{2\hat{j}^{2}_{1}}
				 	\sum_{abcJ}\hJ^{2}\left(\eta^{J}_{c1ab}\Gamma^{J}_{abc2}-\Gamma^{J}_{c1ab}\eta^{J}_{abc2}\right)
					\big(n_{a}n_{b}(1-n_{c})+(1-n_{a})(1-n_{b})n_{c}\big)\,,\\                 
  \label{eq:2bflow}
 \totd{\Gamma^{J}_{1234}}{s}
 				&=\sum_{a}\Big(\left(\eta_{1a}\Gamma^{J}_{a234}-f_{1a}\eta^{J}_{a234} -(-1)^{J-j_{1}-j_{2}} \left[1\leftrightarrow2\right]\right)  - 
						  \left(\eta_{3a}\Gamma^{J}_{12a4}-f_{3a}\eta^{J}_{12a4} -(-1)^{J-j_{3}-j_{4}} \left[3\leftrightarrow4\right]\right)\Big)\notag\\
				&\hphantom{=}+\frac{1}{2}\sum_{ab}\left(\eta^{J}_{12ab}\Gamma^{J}_{ab34}-\Gamma^{J}_{12ab}\eta^{J}_{ab34}\right)\left(1-n_{a}-n_{b}\right)\notag\\
				&\hphantom{=}+\sum_{abJ'}\left(n_{a}-n_{b}\right)
					\left(\sixj{j_{1} & j_{2} & J \\ j_{3} & j_{4} & J'}\left(\overline\eta^{J'}_{1\bar{4}a\bar{b}}\overline\Gamma^{J'}_{a\bar{b}3\bar{2}}-
						  												    \overline\Gamma^{J'}_{1\bar{4}a\bar{b}}\overline\eta^{J'}_{a\bar{b}3\bar{2}}\right) 
																		    -(-1)^{J-j_{1}-j_{2}}\left[1\leftrightarrow2\right]\right)\,,
\end{align}
\end{widetext}
where $\hat{j}=\sqrt{2j+1}$, indices with a bar indicate time-reversed states, and the $\overline{\eta}$ and $\overline{\Gamma}$ matrix elements in the last line of Eq.~ \eqref{eq:2bflow} are obtained by a generalized Pandya transform (see, e.g., \cite{Suhonen:2007wo}),
\begin{equation}
  \overline\OO^{J}_{1\bar{2}3\bar{4}} = - \sum_{J'}\hat{J'}^{2}\sixj{j_{1} & j_{2} & J \\ j_{3} & j_{4} & J'} O^{J'}_{1432}\,.
\end{equation}

\subsection{Choice of Generator}

\begin{figure}[b]
  \includegraphics[width=\columnwidth]{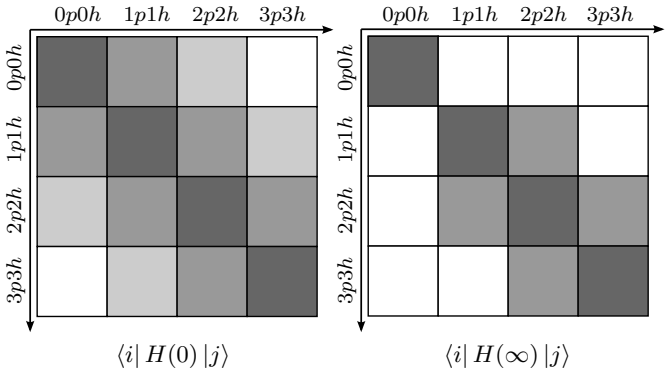}
  \caption{\label{fig:schematic}Schematic representation of the initial and final Hamiltonians, $\HO(0)$ and $\HO(\infty)$, in the many-body Hilbert space spanned by particle-hole excitations of the reference state.}
\end{figure}

\begin{figure*}[t]
  \includegraphics[width=\textwidth]{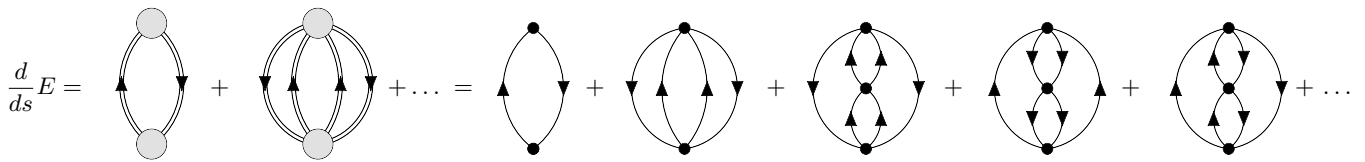}
  \caption{\label{fig:flow}Schematic illustration of the energy flow equation \eqref{eq:0bflow} for the White generator \eqref{eq:eta} in terms of Hugenholtz diagrams (see text). The grey vertices correspond to $\HO(s)$, and the double lines indicate energy denominators calculated with $f(s)$. On the right-hand side, the flow equation is expanded in terms of $\HO(s-\delta s)$ (simple black vertex) and the corresponding energy denominators from $f(s-\delta s)$ (single lines). The dots indicate omitted terms from the expansion of the Epstein-Nesbet energy denominator, as well as additional diagrams generated by the integration step $s-\delta s \to s$.
  }
\end{figure*}

The next step is to identify the ``off-diagonal'' part of the Hamiltonian, which we wish to suppress by integrating the flow equation. To this end, we consider the representation of the flowing Hamiltonian $\HO(s)$, truncated to two-body operators, in a many-body basis consisting of up to $A$-particle-$A$-hole ($ApAh$) excitations of the reference state $\ket{\Phi}$, as sketched in Fig.~\ref{fig:schematic}. The $0p0h$ reference state is coupled to $1p1h$ and $2p2h$ excitations by the matrix elements
\begin{align}
  \matrixe{\Phi}{\HO:\aaO_{p}\aO_{h}:}{\Phi}&=f_{ph}\,,\label{eq:fod}\\
  \matrixe{\Phi}{\HO:\aaO_{p}\aaO_{p'}\aO_{h'}\aO_{h}:}{\Phi}&=\Gamma_{pp'hh'}\label{eq:Gammaod}\,,
\end{align}
and their Hermitian conjugates, which define the one- and two-body pieces of the off-diagonal Hamiltonian $H^{od}$. This $H^{od}$ is now used to construct the generator \cite{White:2002fk,Tsukiyama:2011uq},
\begin{align}
  \etaO
  &=\sum_{ph}\frac{f_{ph}}{\matrixe{ph}{\HO}{ph}-\matrixe{\Phi}{\HO}{\Phi}}:\aaO_{p}\aO_{h}:+\notag\\
  			&\hphantom{=}+\sum_{pp'hh'}\frac{\Gamma_{pp'hh'}}{\matrixe{pp'hh'}{\HO}{pp'hh'}-\matrixe{\Phi}{\HO}{\Phi}}:\aaO_{p}\aaO_{p'}\aO_{h'}\aO_{h}:\notag\\[3pt]
			&\hphantom{=}-\text{H.c.}\notag\\
  &=\sum_{ph}\frac{f_{ph}}{f_{p}-f_{h}-\Gamma_{phph}}:\aaO_{p}\aO_{h}:+\notag\\
  			&\hphantom{=}+\sum_{pp'hh'}\frac{\Gamma_{pp'hh'}}{f_{p}+f_{p'}-f_{h}-f_{h'}+A_{pp'hh'}}:\aaO_{p}\aaO_{p'}\aO_{h'}\aO_{h}:\notag\\[3pt]
			&\hphantom{=}-\text{H.c.}\,,\label{eq:eta}
\end{align}
where $f_{p}=f_{pp}, f_{h}=f_{hh}$, and
\begin{align}
	A_{pp'hh'}&=\Gamma_{pp'pp'}+\Gamma_{hh'hh'}-\Gamma_{phph}\notag\\
			 &\hphantom{=}-\Gamma_{p'h'p'h'}-\Gamma_{ph'ph'}-\Gamma_{p'hp'h}\,.
\end{align}
The energy denominators appearing in Eq.~\eqref{eq:eta} represent a link between the IM-SRG and Many-Body Perturbation Theory (MBPT), which will be discussed in Section \ref{sec:mbpt}. As suggested by White in Ref. \cite{White:2002fk}, we use the Epstein-Nesbet partitioning rather than the more common M{\o}ller-Plesset partitioning \cite{Shavitt:2009} and construct energy denominators using the diagonal matrix elements of $H$ in our chosen matrix representation (Fig.~\ref{fig:schematic}). This naturally regularizes the generator in situations where the difference of the single-particle energies might become small during the IM-SRG flow. 

Because $\eta$ is now given by ratios of energies, $f$ and $\Gamma$ only contribute linearly to the magnitude of the right-hand side of the IM-SRG flow equations \eqref{eq:0bflow}--\eqref{eq:2bflow}. This reduces the stiffness of the flow equations significantly compared to calculations with the canonical Wegner generator, $\eta'=\comm{H}{H^{od}}$ \cite{Wegner:1994dk,Kehrein:2006kx}, where third powers of $f$ and $\Gamma$ appear. 

The use of \eqref{eq:eta} also implies that the off-diagonal matrix elements \eqref{eq:fod}, \eqref{eq:Gammaod} are suppressed at all energy scales simultaneously, i.e.,
\begin{equation}
  \HO^{od}(s) \sim H^{od}(0)e^{-s}\,, 
\end{equation}
in contrast to the canonical Wegner generator, which behaves in a true RG sense and preferentially suppresses matrix elements between states with large energy differences (see the discussion for the free-space SRG in \cite{Bogner:2010pq}). However, since our aim is to solve the many-body problem via the IM-SRG transformation, we are only interested in the limit $s\to\infty$, and then the White generator \eqref{eq:eta} must yield the same result as the Wegner generator, up to truncation errors. We have verified numerically that the energies obtained with these generators differ at the single-$\keV$ level in the present work (also see Ref. \cite{Tsukiyama:2011uq}).

\subsection{\label{sec:mbpt}Connection to Perturbation Theory}
In the previous section, we have already indicated that the White generator \eqref{eq:eta} reveals a clear connection between IM-SRG and MBPT. If we substitute Eq.~ \eqref{eq:eta} into the flow equation for the ground-state energy, Eq.~\eqref{eq:0bflow}, we obtain
\begin{align}
  \totd{E}{s}&=2\sum_{ph}\hat{j}_{p}^{2}\frac{|f_{ph}|^{2}}{f_{p}-f_{h}}\delta_{j_{p}j_{h}}\notag\\
				&\hphantom{=}+\sum_{pp'hh'}\hJ^{2}\frac{|\Gamma^{J}_{pp'hh'}|^{2}}{f_{p}+f_{p'}-f_{h}-f_{h'}} + O(\Gamma^{3}) \label{eq:flowpt}\,,
\end{align}
where we have expanded the Epstein-Nesbet energy denominator as a Neumann series to make the connection with the more commonly used M{\o}ller-Plesset partitioning clearer \cite{Shavitt:2009}. Equation \eqref{eq:flowpt} clearly has the structure of the second-order MBPT correction to the ground-state energy, but note that $f$ and $\Gamma$ both depend on the flow parameter $s$. Thus, the ground-state energy is RG-improved by including contributions from higher-order diagrams into $E(s)$. 

The effect of integrating the energy flow equation \eqref{eq:0bflow} by a single step $s-\delta s \to s$ is illustrated schematically in Fig.~\ref{fig:flow}. Second-order terms in $\HO(s)$, which are added to $E(s)$, contain contributions from second- and third-order diagrams when expressed in $\HO(s-\delta s)$. Note that all topologies for third-order diagrams are generated: the second line of the two-body flow equation \eqref{eq:2bflow} adds particle-particle and hole-hole ladder contributions of $\Gamma(s-\delta s)$ to $\Gamma(s)$, while the third line adds particle-hole contributions, and upon insertion of $\Gamma(s)$ into the energy flow equation, we obtain the corresponding Hugenholtz energy diagrams (see, e.g., \cite{Negele:1998ve,Shavitt:2009}) in Fig.~\ref{fig:flow}. Thus, integrating the IM-SRG flow equations from $s=0$ to $\infty$ corresponds to performing a partial re-summation of the MBPT series. This includes energy corrections for the initial normal-ordered Hamiltonian \eqref{eq:Hno} which are (at least) complete to third order, as well as ladder, ring, and classes of ladder-ring-interference diagrams to all orders. A detailed diagrammatic analysis of the IM-SRG will be presented in a forthcoming publication \cite{Tsukiyama:2012}.

\section{\label{sec:ham_implement}Hamiltonians and Implementation}
In this work, we consider two classes of Hamiltonians. The first one, referred to as $NN+3N$-induced, is based on a chiral $NN$ interaction at next-to-next-to-next-to-leading order (N$^{3}$LO) by Entem and Machleidt, with cutoff $\Lambda=500\,\MeV$  \cite{Entem:2003th}. The resolution scale of this Hamiltonian is subsequently lowered by means of a free-space SRG evolution \cite{Jurgenson:2009bs,Jurgenson:2011zr,Roth:2011kx}, and the induced $3N$ interactions are kept explicitly. 

\begin{figure*}[t]
  \includegraphics[width=\textwidth]{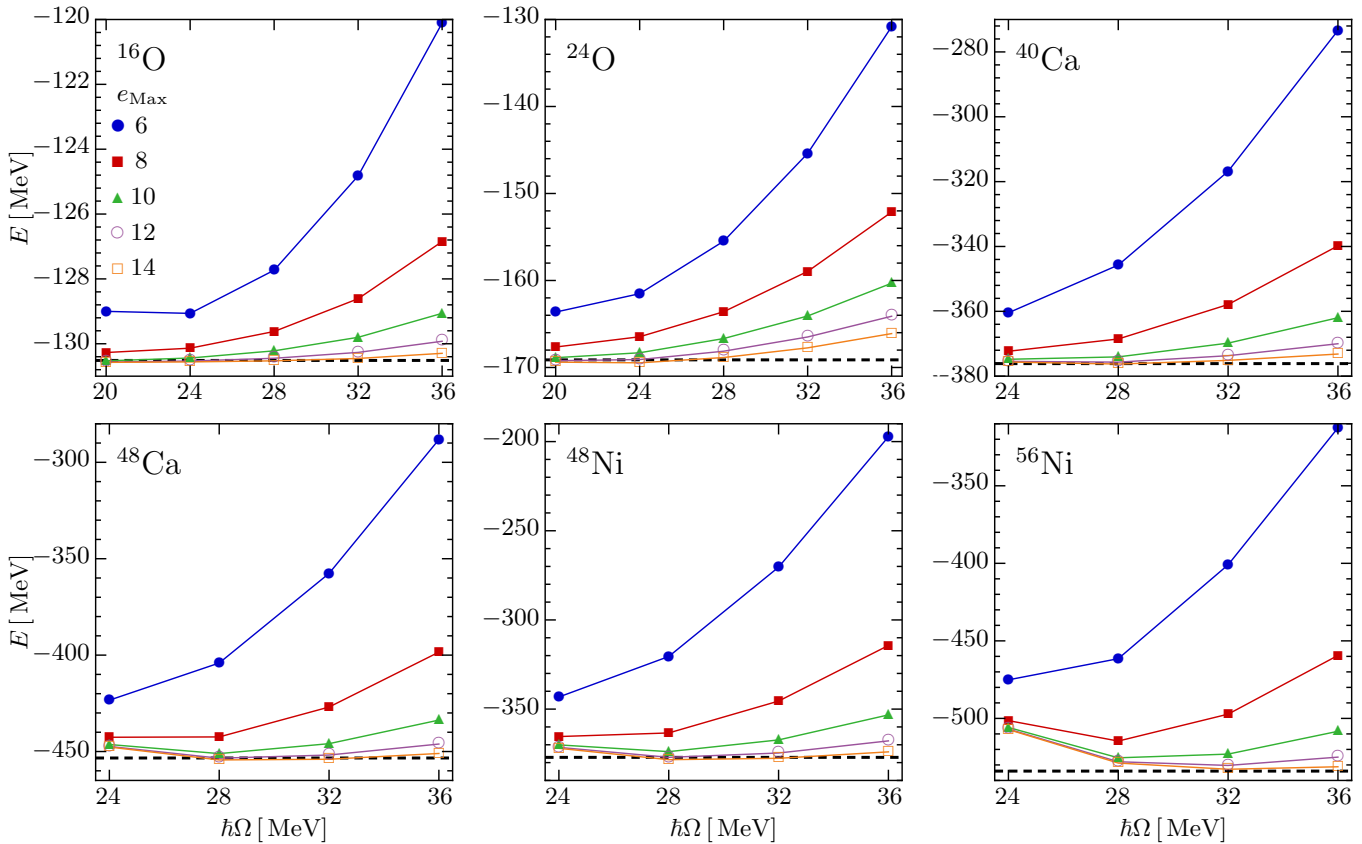}
  \caption{\label{fig:3Nfull_s0625}(Color online) Convergence of the IM-SRG(2) ground-state energy of closed-shell nuclei for the $NN+3N$-full Hamiltonian (see text), evolved to $\lambda=2.0\,\fm^{-1}$. The energies indicated by dashed lines are obtained with the extrapolation method proposed in \cite{Furnstahl:2012ys}.}
\end{figure*}

The second class of Hamiltonians is referred to as $NN+3N$-full in the following, and supplements the aforementioned initial $NN$ Hamiltonian with a local chiral $3N$ interaction at next-to-next-to-leading order (N$^{2}$LO) \cite{Navratil:2007ve}. For a cutoff $\Lambda=500\,\MeV$, this Hamiltonian performs well in calculations for $s$- and $p$-shell nuclei \cite{Gazit:2009qf,Navratil:2009hc,Roth:2011kx}, but when it is evolved to lower resolution scale in order to improve convergence for calculations in the upper $p$-shell and beyond, substantial $4N$ interactions are induced \cite{Roth:2011kx,Roth:2012qf}. Following Ref. \cite{Roth:2012qf}, we reduce the cutoff of the initial $3N$ force to $\Lambda=400\,\MeV$, and use the low-energy constants $c_{D}=-0.2$ and $c_{E}=0.098$ that have been refitted to the $\nuc{He}{4}$ binding energy and triton beta-decay.

The SRG evolution of both types of Hamiltonians is carried out in two- and three-body Jacobi harmonic oscillator (HO) bases with large but finite dimension \cite{Jurgenson:2011zr,Roth:2012vn}. One has to ensure that this model space captures all relevant contributions of the initial Hamiltonian. For small HO frequencies, in particular, the truncation of the model space at fixed HO excitation energy, up to $40\hbar\Omega$ in the present study, leads to additional uncertainties (see \cite{Roth:2012vn} for a detailed discussion). For this reason, the smallest oscillator parameters we use in the following are $\hbar\Omega=20\,\MeV$ in the oxygen isotopes, and $\hbar\Omega=24\,\MeV$ in heavier nuclei. This can be avoided in future applications by either improving the SRG model space in the HO basis \cite{Roth:2012vn} or by carrying out the $3N$ evolution in momentum space \cite{Hebeler:2012ly}.

The SRG-evolved $NN+3N$ Hamiltonian is subsequently transformed to an angular-momentum coupled basis built from single-particle spherical HO states with quantum numbers $e=2n+l\leq\eMax$. For the $3N$ Hamiltonian, we introduce an additional cut $e_{1}+e_{2}+e_{3}\leq \EMax < 3\eMax $ to manage the storage requirements of the matrix elements. Throughout this work, we use $\EMax=14$. As shown in Ref. \cite{Binder:2012uq}, $pf$-shell nuclei are currently the upper limit of \emph{ab initio} calculations with $3N$ interactions, because interactions between nucleons in higher shells are strongly affected or even entirely removed by the $\EMax$ cut. In actual IM-SRG(2) calculations, we find that increasing $\EMax$ from 10 to 12 and from 12 to 14 changes the ground-state energies by $2\%$ and $1\%$, respectively, in the calcium isotopes, and less so in lighter nuclei, for resolution scales $\lambda=1.88,\ldots,2.24\,\fm^{-1}$. For larger $\lambda$, the resulting uncertainty of the ground-state energies grows significantly.

To obtain reference states, we solve the Hartree-Fock equations using the code described in \cite{Hergert:2009zn}, which has been extended to include $3N$ forces. The full three-body Hamiltonian is used in our calculations at this stage. Once a converged HF ground-state is obtained, the nuclear Hamiltonian is normal-ordered w.r.t. to this solution, and the resulting in-medium zero-, one-, and two-body operators serve as the initial values for the IM-SRG flow equations. As discussed above, the residual three-body term is neglected. The flow equations are subsequently integrated using the solver CVODE \cite{Hindmarsh:2005kl} until decoupling is achieved. To check the convergence, we calculate the energy correction from second-order MBPT for the flowing Hamiltonian $\HO(s)$, which is entirely due to the off-diagonal part of the Hamiltonian as defined in Eqs.~\eqref{eq:fod} and \eqref{eq:Gammaod}. We assume that sufficient decoupling is achieved once the perturbative contribution is smaller than  $10^{-6}\,\MeV$, corresponding to relative changes in the flowing ground-state energy of $10^{-7}$ or less in medium-mass nuclei.

To illustrate the convergence of our calculations, we show IM-SRG(2) ground-state energies as a function of the model space size $\eMax$ and the oscillator parameter $\hbar\Omega$ for the SRG-evolved $NN+3N$-full Hamiltonian at $\lambda=2.0\,\fm^{-1}$ in Fig.~\ref{fig:3Nfull_s0625}. The rapid convergence of these ground-state energies is also representative for the $NN+3N$-induced Hamiltonian in the range of studied resolution scales $\lambda=1.88,\ldots,2.24\,\fm^{-1}$. The dashed lines in Fig.~\ref{fig:3Nfull_s0625} indicate energies that were obtained with an extrapolation to infinite HO spaces as proposed in \cite{Furnstahl:2012ys}. Note that the $pf$-shell nuclei, in particular $\nuc{Ni}{56}$, exhibit kinks in the energies around $\hbar\Omega=24\,\MeV$ which are caused by the free-space SRG model space truncation discussed above. We will come back to this subject when we summarize our ground-state energies at the end of Sect.~\ref{sec:results}.

\section{Results\label{sec:results}}
\begin{figure}
  \includegraphics[width=\columnwidth]{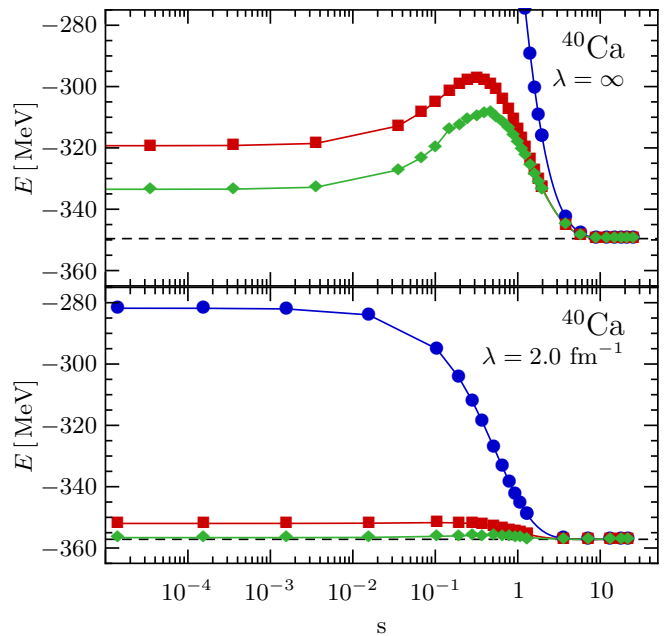}
  \caption{\label{fig:mbpt}
   (Color online) IM-SRG(2) ground-state energy of $\nuc{Ca}{40}$ as a function of the flow parameter $s$ for the bare $NN$ Hamiltonian (top) and the $NN+3N$-induced Hamiltonian with $\lambda=2.0\,\fm^{-1}$ (bottom). Shown are $E$(s) (\symbolcircle[FGBlue]), and $E(s)$ plus second (\symbolbox[FGRed]) and third-order (\symboldiamond[FGGreen]) MBPT corrections for $\HO(s)$. The dashed line indicates the final value $E(\infty)$. All calculations were done with $\eMax=10$, at optimal $\hbar\Omega=32\,\MeV$ (top) and $24\,\MeV$ (bottom). For $s>1.5$ we only show every tenth numerical data point to reduce clutter.
  }
\end{figure}

In this section, we discuss IM-SRG results for the ground-state energies of closed-shell nuclei. We first show how the IM-SRG re-sums correlations into the ground-state energy as we integrate the flow equations. Next, we compare IM-SRG results to other non-perturbative methods to assess its capabilities. Trends of the ground-state energies under the variation of the resolution scale $\lambda$ help us to identify and quantify the uncertainties of our results. We conclude the discussion by giving an overview of IM-SRG ground-state energies for closed-shell nuclei up to $\nuc{Ni}{56}$.

Let us now demonstrate how the IM-SRG re-sums the MBPT series, as explained in Section \ref{sec:mbpt}. To this end, we show the flowing  $\nuc{Ca}{40}$ ground-state energy plus second- and third-order MBPT corrections as a function of the flow parameter $s$ in Fig.~\ref{fig:mbpt}, using both the bare $NN$ Hamiltonian and the $NN+3N$-induced Hamiltonian at a resolution scale $\lambda=2.0\,\fm^{-1}$. Evidently, correlations are summed very rapidly into $E(s)$, and after only 25-30 integration steps ($s\sim 2-3$), the size of MBPT corrections is reduced to less than 1\% of $E(s)$. Note, however, that the initial size as well as the efficiency of the re-summation strongly depends on the initial Hamiltonian. For the bare $NN$ Hamiltonian (see Sect.~\ref{sec:ham_implement}), the reference state is unbound at $E(0)=114\,\MeV$ and only the expansion of single-particle wave-functions in terms of HO states keeps the nuclei from breaking apart. The initial second- and third-order contribution are $-434.1\,\MeV$ and $-14.2\,\MeV$, respectively, but by comparing with the final IM-SRG result $E=-349.6\,\MeV$, we find that the re-summed higher-order contributions are about $-16\,\MeV$, i.e., larger than the third-order term. This indicates the poor convergence properties of the MBPT series. 
For the soft $NN+3N$-induced Hamiltonian with $\lambda=2.0\,\fm^{-1}$ (see Sect.~\ref{sec:ham_implement}), on the other hand, HF + second- and third-order MBPT is a good approximation to the ground-state energy, and the higher-order corrections included via the IM-SRG only amount to about $-500\,\keV$. Note, however, that even for soft Hamiltonians, partial summations of the MBPT series to finite order do not converge systematically, as demonstrated in \cite{Roth:2010ys,Langhammer:2012uq}.

\begin{figure}[t]
  \includegraphics[width=\columnwidth]{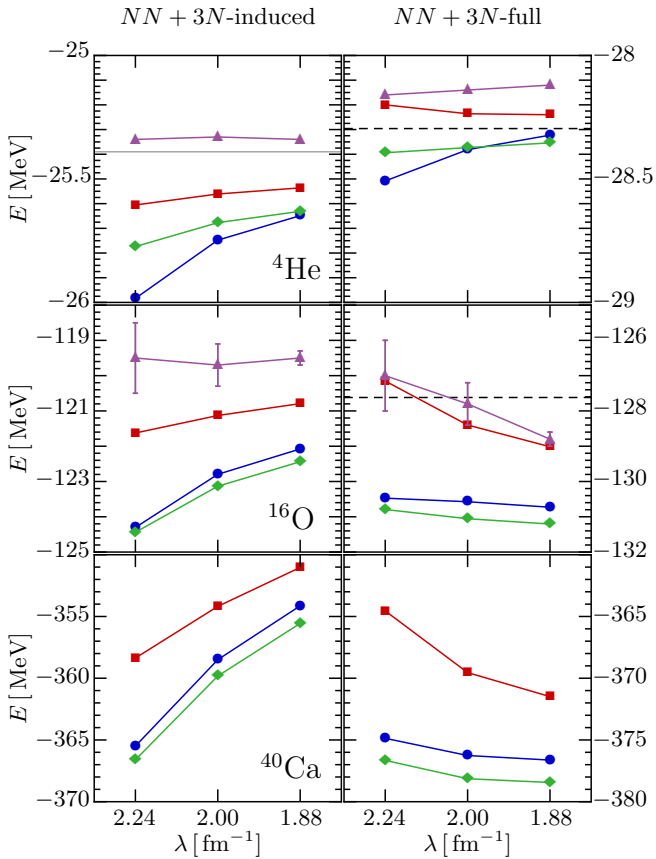}  
  \caption{\label{fig:methods3N}(Color online) Ground-state energies of stable closed-shell nuclei as a function of the resolution scale $\lambda$ for the SRG-evolved $NN + 3N$-induced (left column) and $NN+3N$-full Hamiltonians (right column), using different many-body methods: IM-SRG(2) (\symbolcircle[FGBlue]), CCSD (\symbolbox[FGRed]), $\Lambda$-CCSD(T) (\symboldiamond[FGGreen]), and IT-NCSM (\symboltriangle[FGViolet]). IM-SRG and CC results are obtained with $\eMax=14$, and the IT-NCSM results have been extrapolated to infinite model space, with error bars indicating the uncertainties of  the importance truncation and the model space extrapolation (for $\nuc{He}{4}$, the error bars are smaller than the symbols). The gray solid line for $\nuc{He}{4}$ in the top left panel is the result of a converged NCSM calculation with the bare $NN$ Hamiltonian, $E=-25.39\,\MeV$ (see e.g. \cite{Jurgenson:2009bs}). Dashed black lines are experimental values from \cite{Audi:2002af}.} 
\end{figure}

\begin{figure}
  \includegraphics[width=\columnwidth]{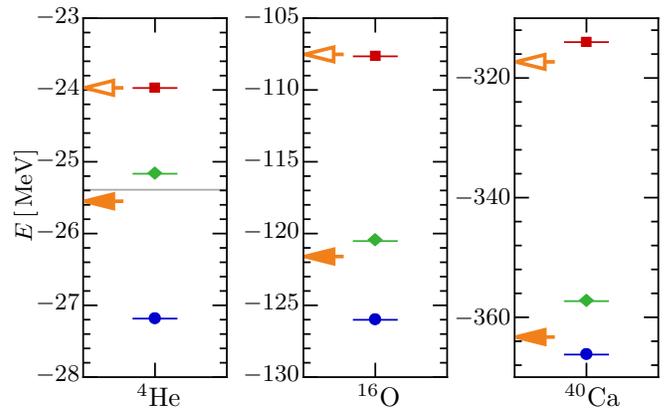}
  \caption{\label{fig:methodsNN}(Color online) Ground-state energies of stable closed-shell nuclei for the bare NN Hamiltonian ($\lambda=\infty$), using IM-SRG(2) (\symbolcircle[FGBlue]), CCSD (\symbolbox[FGRed]), and $\Lambda$-CCSD(T) (\symboldiamond[FGGreen]). IM-SRG and CC results are obtained with $\eMax=14$. The gray solid line for $\nuc{He}{4}$ indicates the result of a converged NCSM calculation, $E=-25.39\,\MeV$ (see e.g. \cite{Jurgenson:2009bs}). The open and solid orange arrows are converged CCSD and $\Lambda$-CCSD(T) results from \cite{Hagen:2010uq,Hagen:2012oq}.} 
\end{figure}

Next, we compare our IM-SRG(2) results for the ground-state energies of $\nuc{He}{4}, \nuc{O}{16},$ and $\nuc{Ca}{40}$ with other non-perturbative methods. To this end, we consider energies from Coupled Cluster calculations with single and double excitations (CCSD) and $\Lambda$-CCSD(T) to assess contributions from triple excitations \cite{Shavitt:2009,Taube:2008kx,Taube:2008vn}. The energies of both methods are obtained using the NO2B approximation (see Sect.~\ref{sec:formalism}) and $\EMax=14$, i.e., the CC and IM-SRG results are based on exactly the same input matrix elements. As a benchmark, we use energies from the quasi-exact Importance-Truncated No-Core Shell Model (IT-NCSM) \cite{Roth:2007fk,Roth:2009eu}, which contain the complete $3N$ interaction. We note that the $\EMax$ cut is naturally compatible with the model-space truncation of the IT-NCSM, and therefore not a source of additional uncertainties in IT-NCSM calculations. In Fig.~\ref{fig:methods3N}, we show such a comparison for the $NN+3N$-induced and $NN+3N$-full Hamiltonians at various resolution scales $\lambda$. 

Surveying the results, we observe that as $\lambda$ decreases, the differences between the IM-SRG(2), CCSD and $\Lambda$-CCSD(T) energies becomes smaller, and all three methods approach the quasi-exact IT-NCSM results in $\nuc{He}{4}$ and $\nuc{O}{16}$. The only deviation from a fairly uniform pattern occurs for $\nuc{He}{4}$ with the $NN+3N$-full Hamiltonian, where the difference between CCSD and IT-NCSM seems to increase slightly, but here the absolute scales of the energy are quite small. The improving agreement of IM-SRG and CC energies is, of course, a manifestation of how the lowering of the resolution scale improves not only the convergence w.r.t. the model space size, but also the convergence of the many-body expansion itself, as observed in the discussion of Fig.~\ref{fig:mbpt} (also see \cite{Binder:2012uq}).

Another interesting observation is that the IM-SRG(2) and $\Lambda$-CCSD(T) energies are surprisingly close, as previously observed in calculations with SRG-evolved $NN$ Hamiltonians in \cite{Tsukiyama:2011uq}. As discussed in Sect.~\ref{sec:mbpt}, the IM-SRG can be understood as a re-summation of the perturbation series for the ground-state energy, and the same is true for CC. We have argued that the IM-SRG(2) energy contains (at least) the complete third-order expressions, similar to CCSD, while $\Lambda$-CCSD(T) is complete through fourth order \cite{Shavitt:2009}. Based on this assessment, and a superficial similarity between the CC cluster operator and the IM-SRG generator $\eta$, one would have expected IM-SRG(2) and CCSD to give similar results. There are, however, differences between the two approaches: for instance, the effective Hamiltonian is always Hermitian in the IM-SRG, which implies that the method has more in common with unitary CC (UCC) approaches rather than the traditional CC \cite{Taube:2006kl}. UCC converges more rapidly to exact Configuration Interaction results than its traditional CC counterpart at a given truncation level because it contains additional many-body contributions, see, e.g., \cite{Kutzelnigg:1991hc}. The proximity of the IM-SRG(2) and $\Lambda$-CCSD(T) results in our calculations suggests that the same is true for the IM-SRG. 

For the $NN+3N$-induced Hamiltonian, we can extend our systematics to $\lambda\to\infty$, i.e., the initial point of the free-space SRG evolution, by performing calculations using the bare $NN$ Hamiltonian, with no induced $3N$ interactions. The results of such calculations are shown in Fig.~\ref{fig:methodsNN}, along with essentially converged CCSD and $\Lambda$-CCSD(T) results in $\eMax\geq 18$ spaces by Hagen et al. \cite{Hagen:2010uq,Hagen:2012oq}. The difference between these results and our $\eMax=14$ CC results illustrate the slower convergence w.r.t. $\eMax$ for the bare $NN$ Hamiltonian in comparison to the SRG-evolved Hamiltonians (cf. Sect.~\ref{sec:ham_implement}).

The calculations with the bare $NN$ Hamiltonian show that the IM-SRG(2) energy lies below the exact NSCM result for $\nuc{He}{4}$, which is possible because the method is not variational. In fact, the IM-SRG(2) energy lies below the $\Lambda$-CCSD(T) energy, i.e., the deviation from the exact result is larger than for $\Lambda$-CCSD(T). The IM-SRG(2) energies for $\nuc{O}{16}$, and $\nuc{Ca}{40}$ also lie below their $\Lambda$-CCSD(T) counterparts. There are different mechanisms which might explain this observation: The IM-SRG(2) might contain diagrams beyond $\Lambda$-CCSD(T) that lead to an overestimation of the binding energy, or IM-SRG(2) might lack diagrams compared to $\Lambda$-CCSD(T) that cancel other contributions. A detailed diagrammatic analysis of the IM-SRG which should clarify this issue is in progress, and will be a subject of a forthcoming publication \cite{Tsukiyama:2012}.

It might be tempting to consider the differences between IM-SRG and CC energies for $\nuc{He}{4}$ and the exact NCSM result for the bare $NN$ Hamiltonian as a measure of uncertainty for the different many-body methods, because it only depends on $\eMax$ and the many-body truncation. We emphasize that this is not possible, because the quality of the many-body truncation manifestly depends on the resolution scale $\lambda$, as discussed in the context of Fig.~\ref{fig:methods3N} above. While the uncertainty due to the use of truncated methods like IM-SRG and CC is reduced for soft interactions, the omission of induced $4N,\ldots, AN$ interactions from the free-space SRG evolution, the omission of the normal-ordered $3N$ interaction, and the $\EMax$ cut in the present work all introduce additional uncertainties, which are hard to disentangle (also see \cite{Binder:2012uq}).

Let us focus on the $NN+3N$-induced Hamiltonian first. The IT-NCSM ground-state energies for $\nuc{He}{4}$ and $\nuc{O}{16}$ shown in the left column of Fig.~\ref{fig:methods3N} exhibit almost no $\lambda$-dependence in the range from $\lambda=1.88$ to $2.24\,\fm^{-1}$, which indicates that omitted induced $4N,\ldots, AN$ interactions are irrelevant for these light nuclei \cite{Roth:2011kx, Jurgenson:2009bs}. The NO2B approximation, i.e., the omission of the normal-ordered $3N$ term in the initial Hamiltonian (see Sect.~\ref{sec:formalism}), causes an overestimation of the CCSD binding energy by 2\% in $\nuc{He}{4}$, and 1\% in the other nuclei studied here \cite{Binder:2012uq}. This overestimation is independent of $\lambda$. If we assume that this remains true for IM-SRG(2) and $\Lambda$-CCSD(T) as well, the variation of the IM-SRG and CC energies with $\lambda$ must be caused by a combination of the many-body truncation and the $\EMax$ cut. Both of these effects become weaker as $\lambda$ decreases: The energy gain from including additional many-body terms is reduced as the convergence of the many-body expansion is improved, and the artificial energy gain that is caused by the omission of repulsive $3N$ matrix elements beyond the $\EMax$ cut is reduced as well. 

In $\nuc{He}{4}$, the deviation between the IM-SRG(2) and IT-NCSM ranges from 2.5\% at $\lambda=2.24\,\fm^{-1}$ to 1\% at $\lambda=1.88\,\fm^{-1}$. The effect of the $\EMax$ cut is negligible here, causing changes in the energy in the single-$\keV$ range. Thus, the deviation is dominated by the 2\% contribution of the NO2B approximation, which amounts to an overestimation of the exact NCSM energy by about $500\,\keV$. If we take this into account by shifting the IM-SRG and CC results upward, the IM-SRG(2) energies lie within a band of $350\,\keV$ around the NCSM result, which leads us to estimate the uncertainty associated with the truncation to be on the order of 1--1.5\%.

Size-extensivity of the IM-SRG(2) implies that the relative truncation error, defined as the deviation from the exact NCSM result, should be of similar size in heavier nuclei. Adding the uncertainties associated with the NO2B approximation and the $\EMax$ cut, which are both about 1\% (see Sect. \ref{sec:ham_implement}), the total estimated uncertainty for $\nuc{O}{16}$ and $\nuc{Ca}{40}$ is about 3.5\%. For $\nuc{O}{16}$, this estimate agrees very well with the largest deviation between IT-NCSM and IM-SRG(2), which is 3.6\%. We emphasize, however, that these estimates rely on the assumption that the effect of the NO2B approximation is the same for IM-SRG(2) and $\Lambda$-CCSD(T) as for CCSD.

For the $NN+3N$-full Hamiltonian (right column of Fig.~\ref{fig:methods3N}), we face the additional complication that the IT-NCSM energies for $\nuc{He}{4}$ and $\nuc{O}{16}$ are themselves $\lambda$-dependent, which indicates that induced $4N,\ldots,AN$ interactions are no longer irrelevant. The underestimation of the experimental $\nuc{He}{4}$ binding energy in the IT-NCSM calculations with the $NN+3N$-full Hamiltonian is caused by the omitted induced $4N$ interaction (the fit of the  low-energy constants was performed with the bare $3N$ interaction). Since the $NN$ part of the Hamiltonian is unchanged from the $NN+3N$-induced case, we can conclude that $4N,\ldots,AN$ interactions must be generated by the SRG evolution of the initial $3N$ interaction \cite{Roth:2011kx}. 

In $\nuc{He}{4}$ and $\nuc{O}{16}$, the relative differences between the IT-NCSM results and the IM-SRG and CC energies are reduced in comparison to the $NN+3N$-induced calculation, to about 1.5\% and 3\%, respectively. Interestingly, the IM-SRG(2) energies now exhibit a weaker $\lambda$-dependence than in the $NN+3N$-induced calculations. This is consistent with the $\Lambda$-CCSD(T) results (also see \cite{Binder:2012uq}) and can be explained by the influence of the $\EMax$ truncation and the omitted induced $4N,\ldots,AN$ interactions in the case of the $NN+3N$-full Hamiltonian. In $\nuc{He}{4}$, the omitted induced interactions are attractive and their effect on the IT-NCSM energies grows only weakly as $\lambda$ decreases. For $\nuc{O}{16}$, on the other hand, the reduction of the IT-NCSM ground-state energy with decreasing $\lambda$ indicates that the net effect of the omitted interactions is repulsive. As mentioned above, for IM-SRG and CC we are also omitting parts of the the repulsive $3N$ interaction by introducing the $\EMax$ cut. The $\lambda$-dependence of the two effects is reverse: With decreasing $\lambda$, the effect of omitted $4N,\ldots,AN$ interactions increases, but the contribution of matrix elements beyond the $\EMax$ cut decreases. This leads to an apparent reduction of the $\lambda$-dependence compared to the $NN+3N$-induced calculations.

\begin{figure}
  \includegraphics[width=\columnwidth]{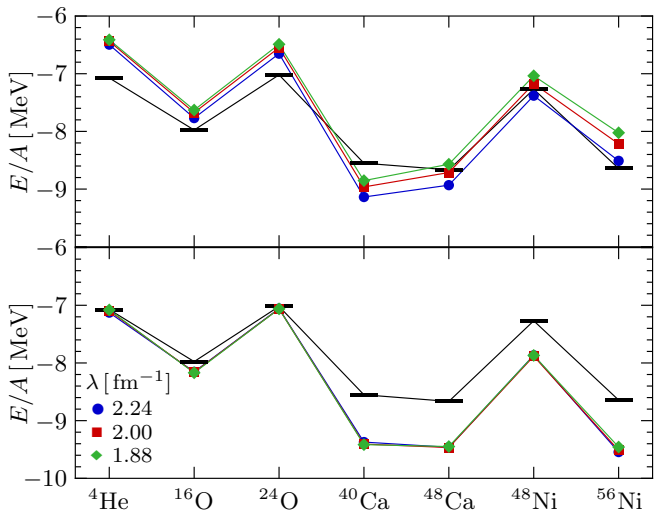}
  \caption{\label{fig:summary}IM-SRG(2) ground-state energy per nucleon of closed-shell nuclei for $NN+3N$-induced (top) and $NN+3N$-full Hamiltonians (bottom) at different resolution scales $\lambda$. Energies are determined at optimal $\hbar\Omega$ for $\eMax=14$. Experimental energies (black bars) are taken from \cite{Audi:2002af}.}
\end{figure}

We conclude the discussion of our results by giving an overview of our IM-SRG(2) ground-state energies in Fig.~\ref{fig:summary}; the corresponding numbers are listed in Table \ref{tab:res3N}. With the $NN+3N$-full Hamiltonian, we obtain a good agreement with experimental binding energies in the $s$- and $p$-shell nuclei, while those of the $sd$- and $pf$-shell nuclei are overestimated by $0.6-0.8\,\MeV$ per nucleon, keeping in mind the previously discussed uncertainties of 3-4\%. A striking feature of Fig.~\ref{fig:summary} is the improvement of the isotopic trends of the ground-state energies with the inclusion of the initial $3N$ force, in particular for the calcium isotopes. Since the effect of a $\lambda$-variation on the ground-state energies is mostly uniform, this improvement appears to be a robust feature. 

For the $NN+3N$-full Hamiltonian, we also find that the aforementioned cancellation which reduces the $\lambda$-dependence of the energies (cf. Fig.~\ref{fig:methods3N}) extends to all nuclei studied in this work. This stability over a wide mass range is quite contrary to the growing $\lambda$-dependence with $A$ which is obtained without the initial $3N$ Hamiltonian. It will be interesting to see if these systematics extend to higher masses, but an extension of our calculations beyond the $pf$-shell will require significant developments in the practical handling of the $3N$ matrix elements. 

\begin{table}
 \begin{tabular*}{\columnwidth}{@{\extracolsep\fill}lcdr@{\hspace{2ex}}dr}
   \hline\hline
   & &
    \multicolumn{2}{c}{$NN+3N$ induced}
   &
    \multicolumn{2}{c}{$NN+3N$ full}
    \\
    \hline
   & &\multicolumn{4}{c}{$E\,[\MeV]$}\\  
   & $\lambda[\fm^{-1}]$ 
   & \multicolumn{1}{c}{$\eMax=14$} 
   & \multicolumn{1}{c}{extrap.} 
   & \multicolumn{1}{c}{$\eMax=14$} 
   & \multicolumn{1}{c}{extrap.} \\
   \hline
   $\nuc{He}{4}$ &  1.88 		& -25.65 	& $-25.65$		& 	-28.32	&	$-28.33$	\\ 
   				&  2.00 		& -25.75 	& $-25.75$		&	-28.38	&	$-28.38$	\\ 
   				&  2.24 		& -25.98 	& $-25.97(1)$	&	-28.51	&	$-28.51$	\\ 
   $\nuc{O}{16}$ & 1.88 		& -122.08 	& $-122.02(7)$	& 	-130.73	& $-130.6(1)$	\\ 
   				&  2.00 		& -122.80 	& $-122.71(9)$	&	-130.57	& $-130.5(1)$	\\ 
   				&  2.24 		& -124.30 	& $-124.1(2)$	& 	-130.47	& $-130.3(2)$	\\ 
   $\nuc{O}{24}$ & 1.88 		& -155.76 	& $-155(1)$	&	-169.49	& $-169(1)$	\\ 
   				&  2.00 		& -157.15 	& $-156.8(9)$	&	-169.47	& $-169(1)$	\\ 
   				&  2.24 		& -159.66 	& $-159.4(6)$	&	-169.36	& $-169.1(9)$	\\ 
   $\nuc{Ca}{40}$ & 1.88 		& -354.14 	& $-354.1(10)$	& 	-376.63	& $-376(1)$	\\ 
   				&  2.00 		& -358.50 	& $-358.5(8)$	& 	-376.26	& $-376.1(8)$	\\ 
   				&  2.24 		& -365.50 	& $-365.7(8)$	& 	-374.87	& $-374.9(7)$	\\ 
   $\nuc{Ca}{48}$ & 1.88 		& -411.34 	& $-409(4)$	& 	-453.55	& $-451(4)$	\\ 
   				&  2.00 		& -418.23 	& $-417(4)$	& 	-454.36	& $-453(5)$	\\ 
   				&  2.24 		& -428.74 	& $-429(5)$	&	-454.17	& $-453(5)$	\\ 
   $\nuc{Ni}{48}$ & 1.88 		& -337.74 	& $-336(3)$	& 	-377.72	& $-375(4)$	\\ 
   				&  2.00 		& -344.30 	& $-343(4)$	& 	-378.36	& $-377(4)$	\\ 
   				&  2.24 		& -354.26 	& $-354(5)$	& 	-377.60	& $-377(5)$	\\ 
   $\nuc{Ni}{56}$ & 1.88 		& -449.35 	& $-453(3)$	& 	-529.62	& $-530(4)$	\\ 
   				&  2.00 		& -460.10 	& $-464(5)$	& 	-532.79	& $-534(4)$	\\ 
   				&  2.24 		& -476.61 	& $-481(7)$	&	-534.48	& $-536(6)$	\\ 
   \hline
   \hline
 \end{tabular*}
 \caption{\label{tab:res3N}Ground-state energies for closed-shell nuclei from IM-SRG(2) calculations with $NN+3N$-induced and $NN+3N$-full Hamiltonians at various values of $\lambda$. Ground-state energies for $\eMax=14$ are taken at the optimal value of $\hbar\Omega$. Extrapolated energies are obtained using the method proposed in \cite{Furnstahl:2012ys}. The uncertainties given in parentheses are solely due to the extrapolation, and determined by the largest difference between the $\eMax=14$ data for all $\hbar\Omega$ and the extrapolated values. Extrapolation uncertainties in $\nuc{He}{4}$ which affect only the third decimal place are not listed. }
\end{table}

Table \ref{tab:res3N} also contains energies which are extrapolated to an infinite HO space, using the extrapolation method proposed in \cite{Furnstahl:2012ys} (also see Fig.~\ref{fig:3Nfull_s0625}). A big advantage of this method is its ability to perform a simultaneous extrapolation for all calculated energies with different $\eMax$ and $\hbar\Omega$, and thereby better constrain the fit model than separate extrapolations at each $\hbar\Omega$. We find the extrapolation works well for the $NN+3N$-induced and $NN+3N$-full Hamiltonians, and all studied nuclei. As an error estimate, we include the maximum value of the scatter of $\eMax=14$ results around the extrapolated values $E_{\infty}$ in Tab.~\ref{tab:res3N}. Since we are limited to $\hbar\Omega\geq 24\,\MeV$ in the calcium isotopes as well as $\nuc{Ni}{48}$, the extrapolation uncertainties are somewhat larger than for the lighter nuclei. Although the $\hbar\Omega=24\,\MeV$ data for the $A=48$ nuclei exhibit signs of the artifact which is caused by the truncation of the free-space SRG model space used for the $3N$ evolution (see Sect.~\ref{sec:ham_implement} and also Fig.~\ref{fig:3Nfull_s0625}), the exclusion of these points from the fit data set does not affect the extrapolation within its uncertainties. For $\nuc{Ni}{56}$, however, the effect is more pronounced, hence we only include data points for $\hbar\Omega\geq28\,\MeV$ when we determine the extrapolated energies in this case.

\section{Conclusions}
We have used the IM-SRG to perform a systematic study of closed-shell nuclei using chiral $NN+3N$ Hamiltonians in the normal-ordered two-body approximation. Our calculations can be carried out in large HO bases of up to 15 major shells due to the polynomial scaling of the IM-SRG with the single-particle basis size. We have demonstrated that the SRG evolution of the initial $NN$ or $NN+3N$ Hamiltonian to a lower resolution scale is under control: Uncertainties caused by the omission of induced $4N,\ldots,AN$ forces are outweighed by the improved convergence behavior of the evolved Hamiltonian in many-body calculations, which reduces truncation and extrapolation uncertainties.

With an initial Hamiltonian consisting of the N$^{3}$LO $NN$ interaction with cutoff $\Lambda=500\,\MeV$, and an N$^{2}$LO $3N$ interaction with a lowered cutoff $\Lambda=400\,\MeV$, we achieve a very good agreement of the IM-SRG(2) ground-state energies with experimental data in light nuclei, while calcium and nickel isotopes are over-bound by $0.6$--$0.8\MeV$ per nucleon, with minimal dependence on the SRG parameter $\lambda$. Our IM-SRG ground-state energies are also in good agreement with results from quasi-exact IT-NCSM, as well as CCSD and $\Lambda$-CCSD(T) calculations \cite{Roth:2011kx,Roth:2012qf,Binder:2012uq}. The similarity of our results to the latter raises interesting questions regarding the diagrammatic content of the IM-SRG method when it is interpreted as a non-perturbative re-summation of the MBPT series. This aspect will be explored in depth in a forthcoming publication \cite{Tsukiyama:2012}. 

Improving on the IM-SRG(2) truncation introduced in this work is straightforward: As a next step, we plan to implement the IM-SRG(3) flow equations, which contain explicit three-body operators. While the computational treatment of $3N$ interactions is expensive and requires additional approximations like the $\EMax$ cut we discussed in the present work, it should be feasible to perform calculations in light nuclei to establish the convergence behavior of the IM-SRG hierarchy of truncations. Effects in heavier nuclei can then be inferred using size-extensivity arguments.

In the near term, we can use the existing technology to study $NN$ and $3N$ Hamiltonian at consistent cutoffs and orders of the chiral expansion, in particular the recently completed N$^{3}$LO interactions \cite{Bernard:2008ql,Bernard:2011ly}. We also plan spectroscopic applications by using the final IM-SRG single-particle energies and residual two-body interaction as input for traditional Shell Model calculations \cite{Tsukiyama:2012fk}. In parallel, we will pursue the generalization and direct application of the IM-SRG to open-shell nuclei via a multi-reference formalism. There are indications from quantum chemistry that a unitary method like the IM-SRG is more robust than traditional CC approaches in such cases \cite{Taube:2006kl,Yanai:2006uq,Yanai:2007kx}. 

\section*{Acknowledgments}
We thank R. Furnstahl, K. Hebeler, K. Tsukiyama, and K. Wendt for useful comments and discussions.

This work is supported in part by the National Science Foundation under Grants No.~PHY-1002478 and PHY-1068648, the NUCLEI SciDAC Collaboration under the U.S. Department of Energy Grants No. DE-SC0008533 and DE-SC0008511, the Deutsche Forschungsgemeinschaft through contract SFB 634, the Helmholtz International Center for FAIR within the framework of the LOEWE program launched by the State of Hesse, the Helmholtz Alliance HA216/EMMI, and the German Federal Ministry of Education and Research (BMBF) through contracts 06DA9040I and 06DA7074I. 

Computing resources were provided by the Ohio Supercomputer Center, the LOEWE-CSC Frankfurt, and the National Energy Research Scientific Computing Center supported by the Office of Science of the U.S. Department of Energy under Contract No. DE-AC02-05CHH11231.

We thank the Institute for Nuclear Theory at the University of Washington for its hospitality and the Department of Energy for partial support during the completion of this work.

\begin{thebibliography}{48}%
\makeatletter
\providecommand \@ifxundefined [1]{%
 \@ifx{#1\undefined}
}%
\providecommand \@ifnum [1]{%
 \ifnum #1\expandafter \@firstoftwo
 \else \expandafter \@secondoftwo
 \fi
}%
\providecommand \@ifx [1]{%
 \ifx #1\expandafter \@firstoftwo
 \else \expandafter \@secondoftwo
 \fi
}%
\providecommand \natexlab [1]{#1}%
\providecommand \enquote  [1]{``#1''}%
\providecommand \bibnamefont  [1]{#1}%
\providecommand \bibfnamefont [1]{#1}%
\providecommand \citenamefont [1]{#1}%
\providecommand \href@noop [0]{\@secondoftwo}%
\providecommand \href [0]{\begingroup \@sanitize@url \@href}%
\providecommand \@href[1]{\@@startlink{#1}\@@href}%
\providecommand \@@href[1]{\endgroup#1\@@endlink}%
\providecommand \@sanitize@url [0]{\catcode `\\12\catcode `\$12\catcode
  `\&12\catcode `\#12\catcode `\^12\catcode `\_12\catcode `\%12\relax}%
\providecommand \@@startlink[1]{}%
\providecommand \@@endlink[0]{}%
\providecommand \url  [0]{\begingroup\@sanitize@url \@url }%
\providecommand \@url [1]{\endgroup\@href {#1}{\urlprefix }}%
\providecommand \urlprefix  [0]{URL }%
\providecommand \Eprint [0]{\href }%
\providecommand \doibase [0]{http://dx.doi.org/}%
\providecommand \selectlanguage [0]{\@gobble}%
\providecommand \bibinfo  [0]{\@secondoftwo}%
\providecommand \bibfield  [0]{\@secondoftwo}%
\providecommand \translation [1]{[#1]}%
\providecommand \BibitemOpen [0]{}%
\providecommand \bibitemStop [0]{}%
\providecommand \bibitemNoStop [0]{.\EOS\space}%
\providecommand \EOS [0]{\spacefactor3000\relax}%
\providecommand \BibitemShut  [1]{\csname bibitem#1\endcsname}%
\let\auto@bib@innerbib\@empty
\bibitem [{\citenamefont {Bogner}\ \emph {et~al.}(2010)\citenamefont {Bogner},
  \citenamefont {Furnstahl},\ and\ \citenamefont {Schwenk}}]{Bogner:2010pq}%
  \BibitemOpen
  \bibfield  {author} {\bibinfo {author} {\bibfnamefont {S.~K.}\ \bibnamefont
  {Bogner}}, \bibinfo {author} {\bibfnamefont {R.~J.}\ \bibnamefont
  {Furnstahl}}, \ and\ \bibinfo {author} {\bibfnamefont {A.}~\bibnamefont
  {Schwenk}},\ }\href {\doibase 10.1016/j.ppnp.2010.03.001} {\bibfield
  {journal} {\bibinfo  {journal} {Prog. Part. Nucl. Phys.}\ }\textbf {\bibinfo
  {volume} {65}},\ \bibinfo {pages} {94} (\bibinfo {year} {2010})} \BibitemShut
  {NoStop}%
%
%
\bibitem [{\citenamefont {Epelbaum}\ \emph {et~al.}(2009)\citenamefont
  {Epelbaum}, \citenamefont {Hammer},\ and\ \citenamefont
  {Mei\ss{}ner}}]{Epelbaum:2009ve}%
  \BibitemOpen
  \bibfield  {author} {\bibinfo {author} {\bibfnamefont {E.}~\bibnamefont
  {Epelbaum}}, \bibinfo {author} {\bibfnamefont {H.-W.}\ \bibnamefont
  {Hammer}}, \ and\ \bibinfo {author} {\bibfnamefont {U.-G.}\ \bibnamefont
  {Mei\ss{}ner}},\ }\href {\doibase 10.1103/RevModPhys.81.1773} {\bibfield
  {journal} {\bibinfo  {journal} {Rev. Mod. Phys.}\ }\textbf {\bibinfo {volume}
  {81}},\ \bibinfo {pages} {1773} (\bibinfo {year} {2009})}\BibitemShut
  {NoStop}%
%
%
\bibitem [{\citenamefont {Bogner}\ \emph {et~al.}(2007)\citenamefont {Bogner},
  \citenamefont {Furnstahl},\ and\ \citenamefont {Perry}}]{Bogner:2007od}%
  \BibitemOpen
  \bibfield  {author} {\bibinfo {author} {\bibfnamefont {S.~K.}\ \bibnamefont
  {Bogner}}, \bibinfo {author} {\bibfnamefont {R.~J.}\ \bibnamefont
  {Furnstahl}}, \ and\ \bibinfo {author} {\bibfnamefont {R.~J.}\ \bibnamefont
  {Perry}},\ }\href {\doibase 10.1103/PhysRevC.75.061001} {\bibfield  {journal}
  {\bibinfo  {journal} {Phys. Rev. C}\ }\textbf {\bibinfo {volume} {75}},\
  \bibinfo {pages} {061001(R)} (\bibinfo {year} {2007})}\BibitemShut
  {NoStop}%
%
%
\bibitem [{\citenamefont {Jurgenson}\ \emph {et~al.}(2009)\citenamefont
  {Jurgenson}, \citenamefont {Navratil},\ and\ \citenamefont
  {Furnstahl}}]{Jurgenson:2009bs}%
  \BibitemOpen
  \bibfield  {author} {\bibinfo {author} {\bibfnamefont {E.~D.}\ \bibnamefont
  {Jurgenson}}, \bibinfo {author} {\bibfnamefont {P.}~\bibnamefont {Navratil}},
  \ and\ \bibinfo {author} {\bibfnamefont {R.~J.}\ \bibnamefont {Furnstahl}},\
  }\href {\doibase 10.1103/PhysRevLett.103.082501} {\bibfield  {journal}
  {\bibinfo  {journal} {Phys. Rev. Lett.}\ }\textbf {\bibinfo {volume} {103}},\
  \bibinfo {pages} {082501} (\bibinfo {year} {2009})}\BibitemShut
  {NoStop}%
%
%
\bibitem [{\citenamefont {Hebeler}(2012)}]{Hebeler:2012ly}%
  \BibitemOpen
  \bibfield  {author} {\bibinfo {author} {\bibfnamefont {K.}~\bibnamefont
  {Hebeler}},\ }\href {\doibase 10.1103/PhysRevC.85.021002} {\bibfield
  {journal} {\bibinfo  {journal} {Phys. Rev. C}\ }\textbf {\bibinfo {volume}
  {85}},\ \bibinfo {pages} {021002} (\bibinfo {year} {2012})}\BibitemShut
  {NoStop}%
%
%
\bibitem [{\citenamefont {Tsukiyama}\ \emph {et~al.}(2011)\citenamefont
  {Tsukiyama}, \citenamefont {Bogner},\ and\ \citenamefont
  {Schwenk}}]{Tsukiyama:2011uq}%
  \BibitemOpen
  \bibfield  {author} {\bibinfo {author} {\bibfnamefont {K.}~\bibnamefont
  {Tsukiyama}}, \bibinfo {author} {\bibfnamefont {S.~K.}\ \bibnamefont
  {Bogner}}, \ and\ \bibinfo {author} {\bibfnamefont {A.}~\bibnamefont
  {Schwenk}},\ }\href {\doibase 10.1103/PhysRevLett.106.222502} {\bibfield
  {journal} {\bibinfo  {journal} {Phys. Rev. Lett.}\ }\textbf {\bibinfo
  {volume} {106}},\ \bibinfo {pages} {222502} (\bibinfo {year}
  {2011})}\BibitemShut {NoStop}%
%
%
\bibitem [{\citenamefont {Tsukiyama}\ \emph {et~al.}(2012)\citenamefont
  {Tsukiyama}, \citenamefont {Bogner},\ and\ \citenamefont
  {Schwenk}}]{Tsukiyama:2012fk}%
  \BibitemOpen
  \bibfield  {author} {\bibinfo {author} {\bibfnamefont {K.}~\bibnamefont
  {Tsukiyama}}, \bibinfo {author} {\bibfnamefont {S.~K.}\ \bibnamefont
  {Bogner}}, \ and\ \bibinfo {author} {\bibfnamefont {A.}~\bibnamefont
  {Schwenk}},\ }\href {\doibase 10.1103/PhysRevC.85.061304} {\bibfield
  {journal} {\bibinfo  {journal} {Phys. Rev. C}\ }\textbf {\bibinfo {volume}
  {85}},\ \bibinfo {pages} {061304} (\bibinfo {year} {2012})}\BibitemShut
  {NoStop}%
%
%
\bibitem [{\citenamefont {Shavitt}\ and\ \citenamefont
  {Bartlett}(2009)}]{Shavitt:2009}%
  \BibitemOpen
  \bibfield  {author} {\bibinfo {author} {\bibfnamefont {I.}~\bibnamefont
  {Shavitt}}\ and\ \bibinfo {author} {\bibfnamefont {R.~J.}\ \bibnamefont
  {Bartlett}},\ }\href@noop {} {\emph {\bibinfo {title} {Many-Body Methods in
  Chemistry and Physics: MBPT and Coupled-Cluster Theory}}}\ (\bibinfo
  {publisher} {Cambridge University Press},\ \bibinfo {year}
  {2009})\BibitemShut {NoStop}%
%
%
\bibitem [{\citenamefont {Roth}\ and\ \citenamefont
  {Navratil}(2007)}]{Roth:2007fk}%
  \BibitemOpen
  \bibfield  {author} {\bibinfo {author} {\bibfnamefont {R.}~\bibnamefont
  {Roth}}\ and\ \bibinfo {author} {\bibfnamefont {P.}~\bibnamefont
  {Navratil}},\ }\href@noop {} {\bibfield  {journal} {\bibinfo  {journal}
  {Phys. Rev. Lett.}\ }\textbf {\bibinfo {volume} {99}},\ \bibinfo {pages}
  {092501} (\bibinfo {year} {2007})}
  \BibitemShut {NoStop}%
%
%
\bibitem [{\citenamefont {Roth}(2009)}]{Roth:2009eu}%
  \BibitemOpen
  \bibfield  {author} {\bibinfo {author} {\bibfnamefont {R.}~\bibnamefont
  {Roth}},\ }\href {\doibase 10.1103/PhysRevC.79.064324} {\bibfield  {journal}
  {\bibinfo  {journal} {Phys. Rev. C}\ }\textbf {\bibinfo {volume} {79}},\
  \bibinfo {pages} {064324} (\bibinfo {year} {2009})}\BibitemShut
  {NoStop}%
%
%
\bibitem [{\citenamefont {Hergert}\ and\ \citenamefont
  {Roth}(2009{\natexlab{a}})}]{Hergert:2009wh}%
  \BibitemOpen
  \bibfield  {author} {\bibinfo {author} {\bibfnamefont {H.}~\bibnamefont
  {Hergert}}\ and\ \bibinfo {author} {\bibfnamefont {R.}~\bibnamefont {Roth}},\
  }\href {\doibase 10.1016/j.physletb.2009.10.100} {\bibfield  {journal}
  {\bibinfo  {journal} {Phys. Lett. B}\ }\textbf {\bibinfo {volume} {682}},\
  \bibinfo {pages} {27} (\bibinfo {year} {2009}{\natexlab{a}})}\BibitemShut
  {NoStop}%
%
%
\bibitem [{\citenamefont {Roth}\ \emph
  {et~al.}(2012{\natexlab{a}})\citenamefont {Roth}, \citenamefont {Binder},
  \citenamefont {Vobig}, \citenamefont {Calci}, \citenamefont {Langhammer},\
  and\ \citenamefont {Navr\'atil}}]{Roth:2012qf}%
  \BibitemOpen
  \bibfield  {author} {\bibinfo {author} {\bibfnamefont {R.}~\bibnamefont
  {Roth}}, \bibinfo {author} {\bibfnamefont {S.}~\bibnamefont {Binder}},
  \bibinfo {author} {\bibfnamefont {K.}~\bibnamefont {Vobig}}, \bibinfo
  {author} {\bibfnamefont {A.}~\bibnamefont {Calci}}, \bibinfo {author}
  {\bibfnamefont {J.}~\bibnamefont {Langhammer}}, \ and\ \bibinfo {author}
  {\bibfnamefont {P.}~\bibnamefont {Navr\'atil}},\ }\href {\doibase
  10.1103/PhysRevLett.109.052501} {\bibfield  {journal} {\bibinfo  {journal}
  {Phys. Rev. Lett.}\ }\textbf {\bibinfo {volume} {109}},\ \bibinfo {pages}
  {052501} (\bibinfo {year} {2012}{\natexlab{a}})}\BibitemShut {NoStop}%
%
%
\bibitem [{\citenamefont {Binder}\ \emph {et~al.}(2012)\citenamefont {Binder},
  \citenamefont {Langhammer}, \citenamefont {Calci}, \citenamefont
  {Navr\'atil},\ and\ \citenamefont {Roth}}]{Binder:2012uq}%
  \BibitemOpen
  \bibfield  {author} {\bibinfo {author} {\bibfnamefont {S.}~\bibnamefont
  {Binder}}, \bibinfo {author} {\bibfnamefont {J.}~\bibnamefont {Langhammer}},
  \bibinfo {author} {\bibfnamefont {A.}~\bibnamefont {Calci}}, \bibinfo
  {author} {\bibfnamefont {P.}~\bibnamefont {Navr\'atil}}, \ and\ \bibinfo
  {author} {\bibfnamefont {R.}~\bibnamefont {Roth}},\ }\href@noop {} {\
  (\bibinfo {year} {2012})},\ \Eprint {http://arxiv.org/abs/1211.4748}
  {arXiv:1211.4748 [nucl-th]} \BibitemShut {NoStop}%
%
%
\bibitem [{\citenamefont {Hagen}\ \emph {et~al.}(2007)\citenamefont {Hagen},
  \citenamefont {Papenbrock}, \citenamefont {Dean}, \citenamefont {Schwenk},
  \citenamefont {Nogga}, \citenamefont {W\l{}och},\ and\ \citenamefont
  {Piecuch}}]{Hagen:2007zc}%
  \BibitemOpen
  \bibfield  {author} {\bibinfo {author} {\bibfnamefont {G.}~\bibnamefont
  {Hagen}}, \bibinfo {author} {\bibfnamefont {T.}~\bibnamefont {Papenbrock}},
  \bibinfo {author} {\bibfnamefont {D.~J.}\ \bibnamefont {Dean}}, \bibinfo
  {author} {\bibfnamefont {A.}~\bibnamefont {Schwenk}}, \bibinfo {author}
  {\bibfnamefont {A.}~\bibnamefont {Nogga}}, \bibinfo {author} {\bibfnamefont
  {M.}~\bibnamefont {W\l{}och}}, \ and\ \bibinfo {author} {\bibfnamefont
  {P.}~\bibnamefont {Piecuch}},\ }\href {\doibase 10.1103/PhysRevC.76.034302}
  {\bibfield  {journal} {\bibinfo  {journal} {Phys. Rev. C}\ }\textbf {\bibinfo
  {volume} {76}},\ \bibinfo {pages} {034302} (\bibinfo {year}
  {2007})}\BibitemShut {NoStop}%
%
%
\bibitem [{\citenamefont {Weinberg}(1996)}]{Weinberg:1996uf}%
  \BibitemOpen
  \bibfield  {author} {\bibinfo {author} {\bibfnamefont {S.}~\bibnamefont
  {Weinberg}},\ }\href@noop {} {\emph {\bibinfo {title} {The Quantum Theory of
  Fields, Vol. I. Foundations}}},\ \bibinfo {edition} {2nd}\ ed.\ (\bibinfo
  {publisher} {Cambridge University Press},\ \bibinfo {address} {UK},\ \bibinfo
  {year} {1996})\BibitemShut {NoStop}%
%
%
\bibitem [{\citenamefont {Brandow}(1967)}]{Brandow:1967tg}%
  \BibitemOpen
  \bibfield  {author} {\bibinfo {author} {\bibfnamefont {B.~H.}\ \bibnamefont
  {Brandow}},\ }\href {\doibase 10.1103/RevModPhys.39.771} {\bibfield
  {journal} {\bibinfo  {journal} {Rev. Mod. Phys.}\ }\textbf {\bibinfo {volume}
  {39}},\ \bibinfo {pages} {771} (\bibinfo {year} {1967})}\BibitemShut
  {NoStop}%
%
%
\bibitem [{\citenamefont {Bartlett}(1981)}]{Bartlett:1981zr}%
  \BibitemOpen
  \bibfield  {author} {\bibinfo {author} {\bibfnamefont {R.~J.}\ \bibnamefont
  {Bartlett}},\ }\href {\doibase 10.1146/annurev.pc.32.100181.002043}
  {\bibfield  {journal} {\bibinfo  {journal} {Annual Review of Physical
  Chemistry}\ }\textbf {\bibinfo {volume} {32}},\ \bibinfo {pages} {359}
  (\bibinfo {year} {1981})}
  \BibitemShut {NoStop}%
%
%
\bibitem [{\citenamefont {Negele}\ and\ \citenamefont
  {Orland}(1998)}]{Negele:1998ve}%
  \BibitemOpen
  \bibfield  {author} {\bibinfo {author} {\bibfnamefont {J.~W.}\ \bibnamefont
  {Negele}}\ and\ \bibinfo {author} {\bibfnamefont {H.}~\bibnamefont
  {Orland}},\ }\href@noop {} {\emph {\bibinfo {title} {Quantum Many-Particle
  Systems}}},\ Advanced Book Classics\ (\bibinfo  {publisher} {Westview
  Press},\ \bibinfo {year} {1998})\BibitemShut {NoStop}%
%
%
\bibitem [{\citenamefont {Navratil}\ \emph {et~al.}(2000)\citenamefont
  {Navratil}, \citenamefont {Vary},\ and\ \citenamefont
  {Barrett}}]{Navratil:2000nm}%
  \BibitemOpen
  \bibfield  {author} {\bibinfo {author} {\bibfnamefont {P.}~\bibnamefont
  {Navratil}}, \bibinfo {author} {\bibfnamefont {J.~P.}\ \bibnamefont {Vary}},
  \ and\ \bibinfo {author} {\bibfnamefont {B.~R.}\ \bibnamefont {Barrett}},\
  }\href@noop {} {\bibfield  {journal} {\bibinfo  {journal} {Phys. Rev.}\
  }\textbf {\bibinfo {volume} {C62}},\ \bibinfo {pages} {054311} (\bibinfo
  {year} {2000})}\BibitemShut {NoStop}%
%
%
\bibitem [{\citenamefont {Navr{\'a}til}\ \emph {et~al.}(2009)\citenamefont
  {Navr{\'a}til}, \citenamefont {Quaglioni}, \citenamefont {Stetcu},\ and\
  \citenamefont {Barrett}}]{Navratil:2009hc}%
  \BibitemOpen
  \bibfield  {author} {\bibinfo {author} {\bibfnamefont {P.}~\bibnamefont
  {Navr{\'a}til}}, \bibinfo {author} {\bibfnamefont {S.}~\bibnamefont
  {Quaglioni}}, \bibinfo {author} {\bibfnamefont {I.}~\bibnamefont {Stetcu}}, \
  and\ \bibinfo {author} {\bibfnamefont {B.~R.}\ \bibnamefont {Barrett}},\
  }\href {http://stacks.iop.org/0954-3899/36/i=8/a=083101} {\bibfield
  {journal} {\bibinfo  {journal} {J. Phys. G}\ }\textbf {\bibinfo {volume}
  {36}},\ \bibinfo {pages} {083101} (\bibinfo {year} {2009})}\BibitemShut
  {NoStop}%
%
%
\bibitem [{\citenamefont {Suhonen}(2007)}]{Suhonen:2007wo}%
  \BibitemOpen
  \bibfield  {author} {\bibinfo {author} {\bibfnamefont {J.}~\bibnamefont
  {Suhonen}},\ }\href@noop {} {\emph {\bibinfo {title} {From Nucleons to
  Nucleus. Concepts of Microscopic Nuclear Theory}}},\ \bibinfo {edition}
  {1st}\ ed.\ (\bibinfo  {publisher} {Springer},\ \bibinfo {address} {Berlin},\
  \bibinfo {year} {2007})\BibitemShut {NoStop}%
%
%
\bibitem [{\citenamefont {White}(2002)}]{White:2002fk}%
  \BibitemOpen
  \bibfield  {author} {\bibinfo {author} {\bibfnamefont {S.~R.}\ \bibnamefont
  {White}},\ }\href {\doibase DOI:10.1063/1.1508370} {\bibfield  {journal}
  {\bibinfo  {journal} {J. Chem. Phys.}\ }\textbf {\bibinfo {volume} {117}},\
  \bibinfo {pages} {7472} (\bibinfo {year} {2002})}\BibitemShut {NoStop}%
%
%
\bibitem [{\citenamefont {Wegner}(1994)}]{Wegner:1994dk}%
  \BibitemOpen
  \bibfield  {author} {\bibinfo {author} {\bibfnamefont {F.}~\bibnamefont
  {Wegner}},\ }\href@noop {} {\bibfield  {journal} {\bibinfo  {journal} {Ann.
  Phys. (Leipzig)}\ }\textbf {\bibinfo {volume} {3}},\ \bibinfo {pages} {77}
  (\bibinfo {year} {1994})}\BibitemShut {NoStop}%
%
%
\bibitem [{\citenamefont {Kehrein}(2006)}]{Kehrein:2006kx}%
  \BibitemOpen
  \bibfield  {author} {\bibinfo {author} {\bibfnamefont {S.}~\bibnamefont
  {Kehrein}},\ }\href@noop {} {\emph {\bibinfo {title} {The Flow Equation
  Approach to Many-Particle Systems}}},\ \bibinfo {series} {Springer Tracts in
  Modern Physics}, Vol.\ \bibinfo {volume} {237}\ (\bibinfo  {publisher}
  {Springer Berlin / Heidelberg},\ \bibinfo {year} {2006})\BibitemShut
  {NoStop}%
%
%
\bibitem [{\citenamefont {Tsukiyama}\ \emph {et~al.}()\citenamefont
  {Tsukiyama}, \citenamefont {Bogner}, \citenamefont {Hergert},\ and\
  \citenamefont {Schwenk}}]{Tsukiyama:2012}%
  \BibitemOpen
  \bibfield  {author} {\bibinfo {author} {\bibfnamefont {K.}~\bibnamefont
  {Tsukiyama}}, \bibinfo {author} {\bibfnamefont {S.~K.}\ \bibnamefont
  {Bogner}}, \bibinfo {author} {\bibfnamefont {H.}~\bibnamefont {Hergert}}, \
  and\ \bibinfo {author} {\bibfnamefont {A.}~\bibnamefont {Schwenk}},\
  }\href@noop {} {}\bibinfo {note} {in preparation}\BibitemShut {NoStop}%
%
%
\bibitem [{\citenamefont {Entem}\ and\ \citenamefont
  {Machleidt}(2003)}]{Entem:2003th}%
  \BibitemOpen
  \bibfield  {author} {\bibinfo {author} {\bibfnamefont {D.~R.}\ \bibnamefont
  {Entem}}\ and\ \bibinfo {author} {\bibfnamefont {R.}~\bibnamefont
  {Machleidt}},\ }\href {http://link.aps.org/doi/10.1103/PhysRevC.68.041001}
  {\bibfield  {journal} {\bibinfo  {journal} {Phys. Rev. C}\ }\textbf {\bibinfo
  {volume} {68}},\ \bibinfo {pages} {041001} (\bibinfo {year}
  {2003})}\BibitemShut {NoStop}%
%
%
\bibitem [{\citenamefont {Jurgenson}\ \emph {et~al.}(2011)\citenamefont
  {Jurgenson}, \citenamefont {Navr\'atil},\ and\ \citenamefont
  {Furnstahl}}]{Jurgenson:2011zr}%
  \BibitemOpen
  \bibfield  {author} {\bibinfo {author} {\bibfnamefont {E.~D.}\ \bibnamefont
  {Jurgenson}}, \bibinfo {author} {\bibfnamefont {P.}~\bibnamefont
  {Navr\'atil}}, \ and\ \bibinfo {author} {\bibfnamefont {R.~J.}\ \bibnamefont
  {Furnstahl}},\ }\href {\doibase 10.1103/PhysRevC.83.034301} {\bibfield
  {journal} {\bibinfo  {journal} {Phys. Rev. C}\ }\textbf {\bibinfo {volume}
  {83}},\ \bibinfo {pages} {034301} (\bibinfo {year} {2011})}\BibitemShut
  {NoStop}%
%
%
\bibitem [{\citenamefont {Roth}\ \emph {et~al.}(2011)\citenamefont {Roth},
  \citenamefont {Langhammer}, \citenamefont {Calci}, \citenamefont {Binder},\
  and\ \citenamefont {Navr\'atil}}]{Roth:2011kx}%
  \BibitemOpen
  \bibfield  {author} {\bibinfo {author} {\bibfnamefont {R.}~\bibnamefont
  {Roth}}, \bibinfo {author} {\bibfnamefont {J.}~\bibnamefont {Langhammer}},
  \bibinfo {author} {\bibfnamefont {A.}~\bibnamefont {Calci}}, \bibinfo
  {author} {\bibfnamefont {S.}~\bibnamefont {Binder}}, \ and\ \bibinfo {author}
  {\bibfnamefont {P.}~\bibnamefont {Navr\'atil}},\ }\href {\doibase
  10.1103/PhysRevLett.107.072501} {\bibfield  {journal} {\bibinfo  {journal}
  {Phys. Rev. Lett.}\ }\textbf {\bibinfo {volume} {107}},\ \bibinfo {pages}
  {072501} (\bibinfo {year} {2011})}\BibitemShut {NoStop}%
%
%
\bibitem [{\citenamefont {Furnstahl}\ \emph {et~al.}(2012)\citenamefont
  {Furnstahl}, \citenamefont {Hagen},\ and\ \citenamefont
  {Papenbrock}}]{Furnstahl:2012ys}%
  \BibitemOpen
  \bibfield  {author} {\bibinfo {author} {\bibfnamefont {R.~J.}\ \bibnamefont
  {Furnstahl}}, \bibinfo {author} {\bibfnamefont {G.}~\bibnamefont {Hagen}}, \
  and\ \bibinfo {author} {\bibfnamefont {T.}~\bibnamefont {Papenbrock}},\
  }\href {\doibase 10.1103/PhysRevC.86.031301} {\bibfield  {journal} {\bibinfo
  {journal} {Phys. Rev. C}\ }\textbf {\bibinfo {volume} {86}},\ \bibinfo
  {pages} {031301} (\bibinfo {year} {2012})}\BibitemShut {NoStop}%
%
%
\bibitem [{\citenamefont {Navr{\'a}til}(2007)}]{Navratil:2007ve}%
  \BibitemOpen
  \bibfield  {author} {\bibinfo {author} {\bibfnamefont {P.}~\bibnamefont
  {Navr{\'a}til}},\ }\href {http://dx.doi.org/10.1007/s00601-007-0193-3}
  {\bibfield  {journal} {\bibinfo  {journal} {Few-Body Systems}\ }\textbf
  {\bibinfo {volume} {41}},\ \bibinfo {pages} {117} (\bibinfo {year} {2007})}
  \BibitemShut {NoStop}%
%
%
\bibitem [{\citenamefont {Gazit}\ \emph {et~al.}(2009)\citenamefont {Gazit},
  \citenamefont {Quaglioni},\ and\ \citenamefont {Navr\'atil}}]{Gazit:2009qf}%
  \BibitemOpen
  \bibfield  {author} {\bibinfo {author} {\bibfnamefont {D.}~\bibnamefont
  {Gazit}}, \bibinfo {author} {\bibfnamefont {S.}~\bibnamefont {Quaglioni}}, \
  and\ \bibinfo {author} {\bibfnamefont {P.}~\bibnamefont {Navr\'atil}},\
  }\href {\doibase 10.1103/PhysRevLett.103.102502} {\bibfield  {journal}
  {\bibinfo  {journal} {Phys. Rev. Lett.}\ }\textbf {\bibinfo {volume} {103}},\
  \bibinfo {pages} {102502} (\bibinfo {year} {2009})}\BibitemShut {NoStop}%
%
%
\bibitem [{\citenamefont {Roth}\ \emph
  {et~al.}(2012{\natexlab{b}})\citenamefont {Roth}, \citenamefont {Binder},
  \citenamefont {Calci},\ and\ \citenamefont {Langhammer}}]{Roth:2012vn}%
  \BibitemOpen
  \bibfield  {author} {\bibinfo {author} {\bibfnamefont {R.}~\bibnamefont
  {Roth}}, \bibinfo {author} {\bibfnamefont {S.}~\bibnamefont {Binder}},
  \bibinfo {author} {\bibfnamefont {A.}~\bibnamefont {Calci}}, \ and\ \bibinfo
  {author} {\bibfnamefont {J.}~\bibnamefont {Langhammer}},\ }\href@noop {} {}
  (\bibinfo {year} {2012}{\natexlab{b}}),\ \bibinfo {note} {in
  preparation}\BibitemShut {NoStop}%
%
%
\bibitem [{\citenamefont {Hergert}\ and\ \citenamefont
  {Roth}(2009{\natexlab{b}})}]{Hergert:2009zn}%
  \BibitemOpen
  \bibfield  {author} {\bibinfo {author} {\bibfnamefont {H.}~\bibnamefont
  {Hergert}}\ and\ \bibinfo {author} {\bibfnamefont {R.}~\bibnamefont {Roth}},\
  }\href {\doibase 10.1103/PhysRevC.80.024312} {\bibfield  {journal} {\bibinfo
  {journal} {Phys. Rev. C}\ }\textbf {\bibinfo {volume} {80}},\ \bibinfo
  {pages} {024312} (\bibinfo {year} {2009}{\natexlab{b}})}\BibitemShut
  {NoStop}%
%
%
\bibitem [{\citenamefont {Hindmarsh}\ \emph {et~al.}(2005)\citenamefont
  {Hindmarsh}, \citenamefont {Brown}, \citenamefont {Grant}, \citenamefont
  {Lee}, \citenamefont {Serban}, \citenamefont {Shumaker},\ and\ \citenamefont
  {Woodward}}]{Hindmarsh:2005kl}%
  \BibitemOpen
  \bibfield  {author} {\bibinfo {author} {\bibfnamefont {A.~C.}\ \bibnamefont
  {Hindmarsh}}, \bibinfo {author} {\bibfnamefont {P.~N.}\ \bibnamefont
  {Brown}}, \bibinfo {author} {\bibfnamefont {K.~E.}\ \bibnamefont {Grant}},
  \bibinfo {author} {\bibfnamefont {S.~L.}\ \bibnamefont {Lee}}, \bibinfo
  {author} {\bibfnamefont {R.}~\bibnamefont {Serban}}, \bibinfo {author}
  {\bibfnamefont {D.~E.}\ \bibnamefont {Shumaker}}, \ and\ \bibinfo {author}
  {\bibfnamefont {C.~S.}\ \bibnamefont {Woodward}},\ }\href {\doibase
  10.1145/1089014.1089020} {\bibfield  {journal} {\bibinfo  {journal} {ACM
  Trans. Math. Softw.}\ }\textbf {\bibinfo {volume} {31}},\ \bibinfo {pages}
  {363} (\bibinfo {year} {2005})}\BibitemShut {NoStop}%
%
%
\bibitem [{\citenamefont {Roth}\ and\ \citenamefont
  {Langhammer}(2010)}]{Roth:2010ys}%
  \BibitemOpen
  \bibfield  {author} {\bibinfo {author} {\bibfnamefont {R.}~\bibnamefont
  {Roth}}\ and\ \bibinfo {author} {\bibfnamefont {J.}~\bibnamefont
  {Langhammer}},\ }\href {\doibase 10.1016/j.physletb.2009.12.046} {\bibfield
  {journal} {\bibinfo  {journal} {Phys. Lett. B}\ }\textbf {\bibinfo
  {volume} {683}},\ \bibinfo {pages} {272 } (\bibinfo {year}
  {2010})}\BibitemShut {NoStop}%
%
%
\bibitem [{\citenamefont {Langhammer}\ \emph {et~al.}(2012)\citenamefont
  {Langhammer}, \citenamefont {Roth},\ and\ \citenamefont
  {Stumpf}}]{Langhammer:2012uq}%
  \BibitemOpen
  \bibfield  {author} {\bibinfo {author} {\bibfnamefont {J.}~\bibnamefont
  {Langhammer}}, \bibinfo {author} {\bibfnamefont {R.}~\bibnamefont {Roth}}, \
  and\ \bibinfo {author} {\bibfnamefont {C.}~\bibnamefont {Stumpf}},\ }\href
  {\doibase 10.1103/PhysRevC.86.054315} {\bibfield  {journal} {\bibinfo
  {journal} {Phys. Rev. C}\ }\textbf {\bibinfo {volume} {86}},\ \bibinfo
  {pages} {054315} (\bibinfo {year} {2012})}\BibitemShut {NoStop}%
%
%
\bibitem [{\citenamefont {Audi}\ \emph {et~al.}(2002)\citenamefont {Audi},
  \citenamefont {Wapstra},\ and\ \citenamefont {Thibault}}]{Audi:2002af}%
  \BibitemOpen
  \bibfield  {author} {\bibinfo {author} {\bibfnamefont {G.}~\bibnamefont
  {Audi}}, \bibinfo {author} {\bibfnamefont {A.~H.}\ \bibnamefont {Wapstra}}, \
  and\ \bibinfo {author} {\bibfnamefont {C.}~\bibnamefont {Thibault}},\ }\href
  {\doibase 10.1016/j.nuclphysa.2003.11.003} {\bibfield  {journal} {\bibinfo
  {journal} {Nucl. Phys. A}\ }\textbf {\bibinfo {volume} {729}},\ \bibinfo
  {pages} {337} (\bibinfo {year} {2002})}\BibitemShut {NoStop}%
%
%
\bibitem [{\citenamefont {Hagen}\ \emph {et~al.}(2010)\citenamefont {Hagen},
  \citenamefont {Papenbrock}, \citenamefont {Dean},\ and\ \citenamefont
  {Hjorth-Jensen}}]{Hagen:2010uq}%
  \BibitemOpen
  \bibfield  {author} {\bibinfo {author} {\bibfnamefont {G.}~\bibnamefont
  {Hagen}}, \bibinfo {author} {\bibfnamefont {T.}~\bibnamefont {Papenbrock}},
  \bibinfo {author} {\bibfnamefont {D.~J.}\ \bibnamefont {Dean}}, \ and\
  \bibinfo {author} {\bibfnamefont {M.}~\bibnamefont {Hjorth-Jensen}},\ }\href
  {\doibase 10.1103/PhysRevC.82.034330} {\bibfield  {journal} {\bibinfo
  {journal} {Phys. Rev. C}\ }\textbf {\bibinfo {volume} {82}},\ \bibinfo
  {pages} {034330} (\bibinfo {year} {2010})}\BibitemShut {NoStop}%
%
%
\bibitem [{\citenamefont {Hagen}(2012)}]{Hagen:2012oq}%
  \BibitemOpen
  \bibfield  {author} {\bibinfo {author} {\bibfnamefont {G.}~\bibnamefont
  {Hagen}}}, \bibinfo{note} {personal communication}\BibitemShut{NoStop}%
%
%
\bibitem [{\citenamefont {Taube}\ and\ \citenamefont
  {Bartlett}(2008{\natexlab{a}})}]{Taube:2008kx}%
  \BibitemOpen
  \bibfield  {author} {\bibinfo {author} {\bibfnamefont {A.~G.}\ \bibnamefont
  {Taube}}\ and\ \bibinfo {author} {\bibfnamefont {R.~J.}\ \bibnamefont
  {Bartlett}},\ }\href {\doibase 10.1063/1.2830236} {\bibfield  {journal}
  {\bibinfo  {journal} {J. Chem. Phys.}\ }\textbf {\bibinfo
  {volume} {128}},\ \bibinfo {eid} {044110} (\bibinfo {year}
  {2008}{\natexlab{a}})}\BibitemShut {NoStop}%
%
%
\bibitem [{\citenamefont {Taube}\ and\ \citenamefont
  {Bartlett}(2008{\natexlab{b}})}]{Taube:2008vn}%
  \BibitemOpen
  \bibfield  {author} {\bibinfo {author} {\bibfnamefont {A.~G.}\ \bibnamefont
  {Taube}}\ and\ \bibinfo {author} {\bibfnamefont {R.~J.}\ \bibnamefont
  {Bartlett}},\ }\href {\doibase 10.1063/1.2830237} {\bibfield  {journal}
  {\bibinfo  {journal} {J. Chem. Phys.}\ }\textbf {\bibinfo
  {volume} {128}},\ \bibinfo {eid} {044111} (\bibinfo {year}
  {2008}{\natexlab{b}})}\BibitemShut {NoStop}%
%
%
\bibitem [{\citenamefont {Taube}\ and\ \citenamefont
  {Bartlett}(2006)}]{Taube:2006kl}%
  \BibitemOpen
  \bibfield  {author} {\bibinfo {author} {\bibfnamefont {A.~G.}\ \bibnamefont
  {Taube}}\ and\ \bibinfo {author} {\bibfnamefont {R.~J.}\ \bibnamefont
  {Bartlett}},\ }\href {\doibase 10.1002/qua.21198} {\bibfield  {journal}
  {\bibinfo  {journal} {Int. J. Quantum Chem.}\ }\textbf
  {\bibinfo {volume} {106}},\ \bibinfo {pages} {3393} (\bibinfo {year}
  {2006})}\BibitemShut {NoStop}%
%
%
\bibitem [{\citenamefont {Kutzelnigg}(1991)}]{Kutzelnigg:1991hc}%
  \BibitemOpen
  \bibfield  {author} {\bibinfo {author} {\bibfnamefont {W.}~\bibnamefont
  {Kutzelnigg}},\ }\href {http://dx.doi.org/10.1007/BF01117418} {\bibfield
  {journal} {\bibinfo  {journal} {Theoret. Chimica Acta)}\ }\textbf {\bibinfo
  {volume} {80}},\ \bibinfo {pages} {349} (\bibinfo {year} {1991})}
  \BibitemShut {NoStop}%
%
%
\bibitem [{\citenamefont {Bernard}\ \emph {et~al.}(2008)\citenamefont
  {Bernard}, \citenamefont {Epelbaum}, \citenamefont {Krebs},\ and\
  \citenamefont {Meissner}}]{Bernard:2008ql}%
  \BibitemOpen
  \bibfield  {author} {\bibinfo {author} {\bibfnamefont {V.}~\bibnamefont
  {Bernard}}, \bibinfo {author} {\bibfnamefont {E.}~\bibnamefont {Epelbaum}},
  \bibinfo {author} {\bibfnamefont {H.}~\bibnamefont {Krebs}}, \ and\ \bibinfo
  {author} {\bibfnamefont {U.-G.}\ \bibnamefont {Meissner}},\ }\href {\doibase
  10.1103/PhysRevC.77.064004} {\bibfield  {journal} {\bibinfo  {journal} {Phys.
  Rev. C}\ }\textbf {\bibinfo {volume} {77}},\ \bibinfo {pages} {064004}
  (\bibinfo {year} {2008})}\BibitemShut {NoStop}%
%
%
\bibitem [{\citenamefont {Bernard}\ \emph {et~al.}(2011)\citenamefont
  {Bernard}, \citenamefont {Epelbaum}, \citenamefont {Krebs},\ and\
  \citenamefont {Mei\ss{}ner}}]{Bernard:2011ly}%
  \BibitemOpen
  \bibfield  {author} {\bibinfo {author} {\bibfnamefont {V.}~\bibnamefont
  {Bernard}}, \bibinfo {author} {\bibfnamefont {E.}~\bibnamefont {Epelbaum}},
  \bibinfo {author} {\bibfnamefont {H.}~\bibnamefont {Krebs}}, \ and\ \bibinfo
  {author} {\bibfnamefont {U.-G.}\ \bibnamefont {Mei\ss{}ner}},\ }\href
  {\doibase 10.1103/PhysRevC.84.054001} {\bibfield  {journal} {\bibinfo
  {journal} {Phys. Rev. C}\ }\textbf {\bibinfo {volume} {84}},\ \bibinfo
  {pages} {054001} (\bibinfo {year} {2011})}\BibitemShut {NoStop}%
%
%
\bibitem [{\citenamefont {Yanai}\ and\ \citenamefont
  {Chan}(2006)}]{Yanai:2006uq}%
  \BibitemOpen
  \bibfield  {author} {\bibinfo {author} {\bibfnamefont {T.}~\bibnamefont
  {Yanai}}\ and\ \bibinfo {author} {\bibfnamefont {G.~K.-L.}\ \bibnamefont
  {Chan}},\ }\href {\doibase 10.1063/1.2196410} {\bibfield  {journal} {\bibinfo
   {journal} {J. Chem. Phys.}\ }\textbf {\bibinfo {volume}
  {124}},\ \bibinfo {eid} {194106} (\bibinfo {year} {2006})}\BibitemShut
  {NoStop}%
%
%
\bibitem [{\citenamefont {Yanai}\ and\ \citenamefont
  {Chan}(2007)}]{Yanai:2007kx}%
  \BibitemOpen
  \bibfield  {author} {\bibinfo {author} {\bibfnamefont {T.}~\bibnamefont
  {Yanai}}\ and\ \bibinfo {author} {\bibfnamefont {G.~K.-L.}\ \bibnamefont
  {Chan}},\ }\href {\doibase 10.1063/1.2761870} {\bibfield  {journal} {\bibinfo
   {journal} {J. Chem. Phys.}\ }\textbf {\bibinfo {volume}
  {127}},\ \bibinfo {eid} {104107} (\bibinfo {year} {2007})}\BibitemShut
  {NoStop}%
\end{thebibliography}
%

\end{document}